\documentclass{aa} 

\usepackage{natbib}

\usepackage{graphicx}
\usepackage{bm}
\usepackage{courier}
\usepackage{txfonts}
\usepackage{hyperref}
\hypersetup{
    colorlinks = true,
    allcolors = blue
}
\usepackage{twoopt}
\makeatletter
  \newcommandtwoopt{\citeads}[3][][]{\href{http://adsabs.harvard.edu/abs/#3}%
    {\def\hyper@linkstart##1##2{}%
     \let\hyper@linkend\@empty\citealp[#1][#2]{#3}}}
  \newcommandtwoopt{\citepads}[3][][]{\href{http://adsabs.harvard.edu/abs/#3}%
    {\def\hyper@linkstart##1##2{}%
     \let\hyper@linkend\@empty\citep[#1][#2]{#3}}}
  \newcommandtwoopt{\citetads}[3][][]{\href{http://adsabs.harvard.edu/abs/#3}%
    {\def\hyper@linkstart##1##2{}%
     \let\hyper@linkend\@empty\citet[#1][#2]{#3}}}
  \newcommandtwoopt{\citeyearads}[3][][]%
    {\href{http://adsabs.harvard.edu/abs/#3}
    {\def\hyper@linkstart##1##2{}%
     \let\hyper@linkend\@empty\citeyear[#1][#2]{#3}}}
\makeatother

\begin{document}

   \title{Chronology of our Galaxy from Gaia colour-magnitude diagram fitting (ChronoGal) - III.  Age and metallicity distribution of Gaia-Sausage-Enceladus stars near the Sun}

   \author{Yllari K. González-Koda\thanks{yllarikoda@gmail.com}\inst{1}
   \and Tomás Ruiz-Lara\inst{1,2}
   \and Carme Gallart\inst{3,4} 
   \and Edoardo Ceccarelli \inst{5,6} 
   \and Emma Dodd \inst{7,13}
   \and Emma Fernández-Alvar \inst{3,4} \and Santi Cassisi\inst{8,9}
   \and Francisco Surot\inst{3} 
   \and Fernando Aguado-Agelet \inst{10,11} 
   \and Davide Massari \inst{5}
   \and Matteo Monelli \inst{3,11,12}
   \and Thomas M. Callingham\inst{7}
   \and Amina Helmi \inst{7}
   \and Guillem Aznar-Menargues\inst{4}
   \and David Mirabal\inst{4}  
   \and Alicia Rivero\inst{3,4} 
   \and Anna B. Queiroz \inst{3,4}
   .
   }

   \institute{Universidad de Granada, Departamento de Física Teórica y del Cosmos, Campus Fuente Nueva, Edificio Mecenas,
18071 Granada, Spain 
             \and  
             Instituto Carlos I de Física Teórica y Computacional, Facultad de Ciencias, E-18071 Granada, Spain
             \and  
             Instituto de Astrofísica de Canarias, Calle Vía Láctea s/n, E-38206 La Laguna, Tenerife, Spain
             \and  
             Departamento de Astrofísica, Universidad de La Laguna, 38205 La Laguna, Tenerife, Spain
             \and  
             INAF – Osservatorio di Astrofisica e Scienza dello Spazio di Bologna, Via Gobetti 93/3, 40129 Bologna, Italy
             \and 
             Dipartimento di Fisica e Astronomia, Università degli Studi di Bologna, Via Piero Gobetti 93/2, 40129 Bologna, Italy
             \and 
             Kapteyn Astronomical Institute, University of Groningen, Landleven 12, 9747 AD Groningen, The Netherlands
             \and 
             INAF – Astronomical Observatory of Abruzzo, via M. Maggini, sn, 64100 Teramo, Italy
             \and 
             INFN, Sezione di Pisa, Largo Pontecorvo 3, 56127 Pisa, Italy
             \and 
             atlanTTic, Universidade de Vigo, Escola de Enxeñaría de Telecomunicación, 36310 Vigo, Spain
             \and 
             Universidad de La Laguna, Avda. Astrofísico Fco. Sánchez, E-38205 La Laguna, Tenerife, Spain
             \and 
             INAF – Osservatorio Astronomico di Roma, Via Frascati 33, 00078 Monte Porzio Catone Roma, Italy
             \and 
             Institute for Computational Cosmology, Department of Physics, Durham University, South Road, Durham DH1 3LE, UK
             \\
             }

   \date{Received XXXX; accepted YYYY}

 
  \abstract
   {Gaia-Sausage-Enceladus is considered the last major merger that contributed to the formation of the Milky Way. Its remnants dominate the nearby accreted stellar halo of the Milky Way.}
   {We aim to characterise the star formation history of Gaia-Sausage-Enceladus
   through the age and metallicity of its stellar populations.}
   {From Gaia DR3 data, we dynamically define 
   three Gaia-Sausage-Enceladus samples with 
   different criteria and possible degrees
   of contamination from other substructures in the
   halo. Then, 
   we derive the stellar age and metallicity distributions using the CMDfit.Gaia package.}
   {
   We identify three main populations of stars and a fourth 
   smaller one following an almost linear age-[M/H] relation. The three oldest populations correspond to the bulk of the star formation that lasted for, at least, 
   $\sim$3-4 Gyr and ended about 10 Gyr ago, its metallicities
   ranging from $-$1.7 to $-$0.8. 
   We categorise these populations into two main 
   epochs: the evolution of GSE in isolation and the 
   merger event. This separation finds independent support from the age-metallicty relation of GSE globular clusters (Aguado-Agelet 
   et al., subm.).
   The fourth population is younger and more metal-rich, at $\sim$8.5 Gyr 
   and [M/H]$\sim-0.4$; its link to GSE is unclear.}
   {}


   \titlerunning{Chronogal III. Age and metallicity distribution of GSE
stars near the Sun}
   \authorrunning{González-Koda, Y. K., et al}
   \maketitle
%

\section{Introduction}

One of the key elements of our current understanding of galaxy formation and evolution under the $\Lambda$CDM paradigm is the gravitational interaction between galaxies. In particular, the spiral galaxies that we can observe today ($z=0$), such as our own, have potentially experienced dozens of mergers~\citep{Abadi2003, BullockJohnston2005, Khoperskov2023}, most of them during the early stages of the Universe. Such accretions involve a gain in mass~\citep[e.g.][]{Moster2020}, can lead to important changes in the galaxy morphology and chemo-dynamical properties traced to the present day~\citep[e.g.][]{Toomre1972, Pearson2019, Buck2023}, and can induce events of increased star formation activity \citep{Tissera2000, RuizLara2020, orkney2022}. 
The stellar halo of a galaxy such as the Milky Way (MW) is mainly composed of debris from past accretions, along with heated in-situ disc stars~\citep{Helmi2020, Deason2024}, offering a unique opportunity to unveil the past merger history of galaxies. 

To properly understand the formation of the MW halo we need to distinguish stellar populations with different origins (formed in situ or linked to different accreted building blocks). Even after being fully disrupted by the MW, accreted stars are expected to be clumped together in the integrals of motion space. Thus, many integrals of motion proxies have been proposed to achieve this separation, such as the energy-momentum space~\citep[$E-L_z-L_\perp$, e.g.][]{Helmi2000} and action-angle coordinates~\citep[e.g., ][]{Feuillet2021, Myeong2019}.
Although it is not yet clear to what extent the dynamical space can be used to fully disentangle the separate accretion events in the MW~\citep[e.g,][]{Jean-Baptiste2017,Pagnini2023}, this approach has been 
one of the main tools used to shape our current understanding of the MW's halo.




The {\it Gaia} Data Release 3~\citep[DR3;][]{Gaia2023} included a large sample of stars with radial velocities obtained by the Radial Velocity Spectrometer (RVS) instrument, which together with the positions and proper motions, results in 6D information that is enabling the systematic selection of stars in the dynamical space. This data provides insights on the different building blocks of the MW that constitute possible past accretion events (e.g.,  ~\citealt{Naidu2020, Lovdal2022, Dodd2023, Sante2024}), most notably, that of the progenitor of Gaia-Sausage-Enceladus~\citep[GSE; ][]{Helmi2018, Belokurov2018}.  
With a stellar mass above $10^8$M${}_\odot$ (some estimates: $6\times 10^8$M${}_\odot$,~\citealt{Helmi2018}, $10^{8.85-9.85}$ $M_\odot$,~\citealt{Feuillet2020},
$1.45\times 10^8$M${}_\odot$,~\citealt{Lane2023}), GSE is believed to correspond to the debris of the last massive accretion of the MW, which would have had a deep impact on its evolution. Indeed, an important merger-induced burst of star formation, both observed \citep{Gallart2019} and predicted by models \citep{Grand2020, orkney2022}, would have led to an important increase of the thick disc mass~\citep[see also][]{Helmi2018}, and the energy of the impact would have heated in-situ stars of the proto-MW into a halo-like configuration \citep{Gallart2019, Belokurov2020_Splash, Grand2020}. It may also have triggered the formation of the bar~\citep{Merrow2024}. 


Many approaches have been proposed to characterise the stellar content of the structures in the halo as means to gain insight into the star formation history (SFH) of their progenitors. Some of these include: i) chemical characterisation~\citep[e.g.,][]{Fernandez-Alvar_2018twoHaloPops, Aguado2021_GES_Sequoia, Matsuno2022I, Matsuno2022II, Horta2023, Ernandes2024, Ceccarelli2024}, but the available spectroscopic data for them is still scarce and the links that can be made with their SFH blurry; ii) isochrone fitting~\citep[e.g.,][]{RuizLara2022, Giribaldi2023}, and iii) colour-magnitude diagram (CMD) fitting based on 5D selections~\citep[][]{RuizLara2022b, Dodd2024}. The latter, CMD fitting, commonly used  to compute SFHs of Local Group dwarf galaxies~\citep[][]{Gallart1999, Tolstoy2009, Cignoni2010}, has started to be applied to {\it Gaia} data~\citep[e.g.,][]{Gallart2019, RuizLara2020, DalTio2021, Mazzi2024, Gallart2024}, allowing for precise calculations of age and metallicity distributions with colours and absolute magnitudes (and corresponding, distance, extinction, and reddening) of the stars as input. In this paper, we will use the latter methodology to derive, for the first time, the distribution of ages and metallicities of nearby halo stellar samples that can be associated kinematically to GSE, using 6D information. This will provide a direct characterization of the evolutionary history of GSE from the fossil record of its stars, shed light onto its orbital history around the MW, and a further restrict its accretion time.

Unfortunately, studies using the RVS sample, or any 6-dimensional characterisation of the MW, are typically affected by the difficulty of characterising the selection and completeness function of the survey. Its modelling is a key requirement to draw a comparison between observational data and theoretical models. This difficulty has been 
overcome recently in the context of the CMD fitting methodology~(Fernandez-Alvar et al., subm.), allowing its application to 6D samples within the ChronoGal project~\citep{Gallart2024}.

In this work, we employ two dynamical selections of GSE stars closely related to the recent works of~\citet[][hereafter, D23]{Dodd2023} and~\citet[][hereafter, H24]{Horta2024}. The former is based on the energy-momentum space $E-L_z-L_\perp$, and the latter on the action-angle coordinates $\sqrt{J_R}-J_\phi$. This way, we aim to cover a range of usual 6D GSE selections to build a complete picture of its stellar content via CMD fitting. 

This article is structured as follows. 
In Sec.~\ref{subsec:Method_Data} we define the halo 
parent sample, GSE selection in 
dynamical space, and quality cuts. Sec.~\ref{subsec:Method_SFH} describes the SFH derivation with CMD fitting using CMDft.\textit{Gaia}. Then, 
in Sec.~\ref{sec:results}, we discuss the results and compare them with the literature, 
drawing a picture of the possible merger scenarios that could have led to the derived 
age-metallicity distribution. The summary and main results can be found in Sec.~\ref{sec:conclusions}. Finally, appendices are dedicated to methodology-related tests discussed throughout the text.

%
\section{Methodology}\label{sec:Method}

In this section we describe the dynamical selections based on 6D 
information, which will result in three GSE samples defined in the energy-angular momentum and action-angle space. We then present the CMD fitting methodology followed to characterise their stellar content in terms of their age and metallicity.

%
\subsection{Data and sample definition}\label{subsec:Method_Data}

We present first a brief summary of the procedure followed in D23 to define the GSE and parent samples used in this work. 
From the $\sim$33 million stars with available radial velocity
in { \it Gaia} DR3, D23 reduced the sample to a local sphere of 2.5 kpc around the Sun, with distances defined as the inverse of the zero-point corrected parallax, $\varpi$, given by~\cite{Lindegren2021}\footnote{Provided in the {\tt gaiadr3-zeropoint} Python package by the {\it Gaia} collaboration.} (and parallax over error greater than 5). Then, additional quality cuts (radial velocity error, \texttt{ruwe}, etc) are imposed to ensure the rigorous position of stars in the $E-L_z-L_\perp$ space.

The positions and velocities (6D information) are assigned taking into account $R_\odot=8.2$ kpc~\citep{McMillan2017}, a local standard of rest $|\bm{V}_\text{LSR}|=238$ kms${}^{-1}$~\citep{McMillan2017}, and $(U,V,W)_\odot=(11.1, 12.24, 7.25)$ km/s~\citep{Schonrich2010}. The halo population was defined by a cut in velocity $|\bm{V}-\bm{V}_\text{LSR}|>210$ km/s, with $\bm{V}$ the velocity of each star. Then, the dynamical coordinates were calculated: the  energy, $E$; momentum in the galactocentric vertical direction, $L_z$; and perpendicular angular momentum, $L_\perp=\sqrt{L_x^2+L_y^2}$ (a proxy for a third integral of motion). The potential is comprised of a Miyamoto-Nagai disk with parameters $(a_d,b_d)=(6.5, 0.26)$ kpc, $M_d=9.3\times 10^{10} M_\odot$; a Hernquist 
bulge with $c_b=0.7$ kpc, $M_b=3.0\times 10^{10} M_\odot$; and
a NFW halo with $r_s=21.5$ kpc, $c_h=12$ and $M_\text{halo}=10^{12} M_\odot$ ~\citep[see][for more details]{Lovdal2022}.

A single-linkage clustering algorithm was applied to find substructures inside the stellar halo, appearing as overdensities in the $E-
L_z-L_\perp$ space~\citep[][]{Lovdal2022, RuizLara2022}. The result was an image of the stellar halo with, tentatively, all overdensities (clusters) labelled and 
categorised in different in-situ and 
accreted origins. 

Finally, since all found clusters may not be independent, they were joined considering the Mahalanaobis distance to each other,
\begin{equation}
    D = \sqrt{(\mu_1 - \mu_2)^T (\Sigma_1 + \Sigma_2)^{-1}(\mu_1 - \mu_2)},
\end{equation}
where $\mu_1, \mu_2$ and $\Sigma_1, \Sigma_2$ are the
means and covariance matrices of pairs of clusters, each effectively approximated to an ellipsoidal distribution. The GSE retrieved sample given by D23 is the result of linking a total of 36 clusters.
From now on, we will call this sample GSE-original.

We define two samples of likely GSE members, obtained by selecting halo stars in a region of the dynamical space similar to that occupied by GSE-original. 
Even if GSE-original by definition carries with it
a number of quality selections, the stars in this
step will be selected without any quality
cut; the necessary cleaning of all samples tailored for the CMD fitting will be done and fully explained 
right after all samples are defined.
The first of these two samples, GSE-cluster, is the result of selecting stars in the halo whose Mahalanaobis distance is under $D=2.13$~\citep[which corresponds to the 80$^{\rm th}$ distance percentile,][]{Lovdal2022} to any of the individual 36 clusters that comprise GSE-original, that is, the shape of each cluster is approximated to an ellipsoid and then joined. The second sample, which we will call GSE-group, is obtained by selecting halo stars with a distance under $D=2.13$ from the mean position of GSE-original, that is, approximating the irregular GSE-original distribution by an ellipsoid.
\begin{figure}
    \centering
    \includegraphics[width=\linewidth]{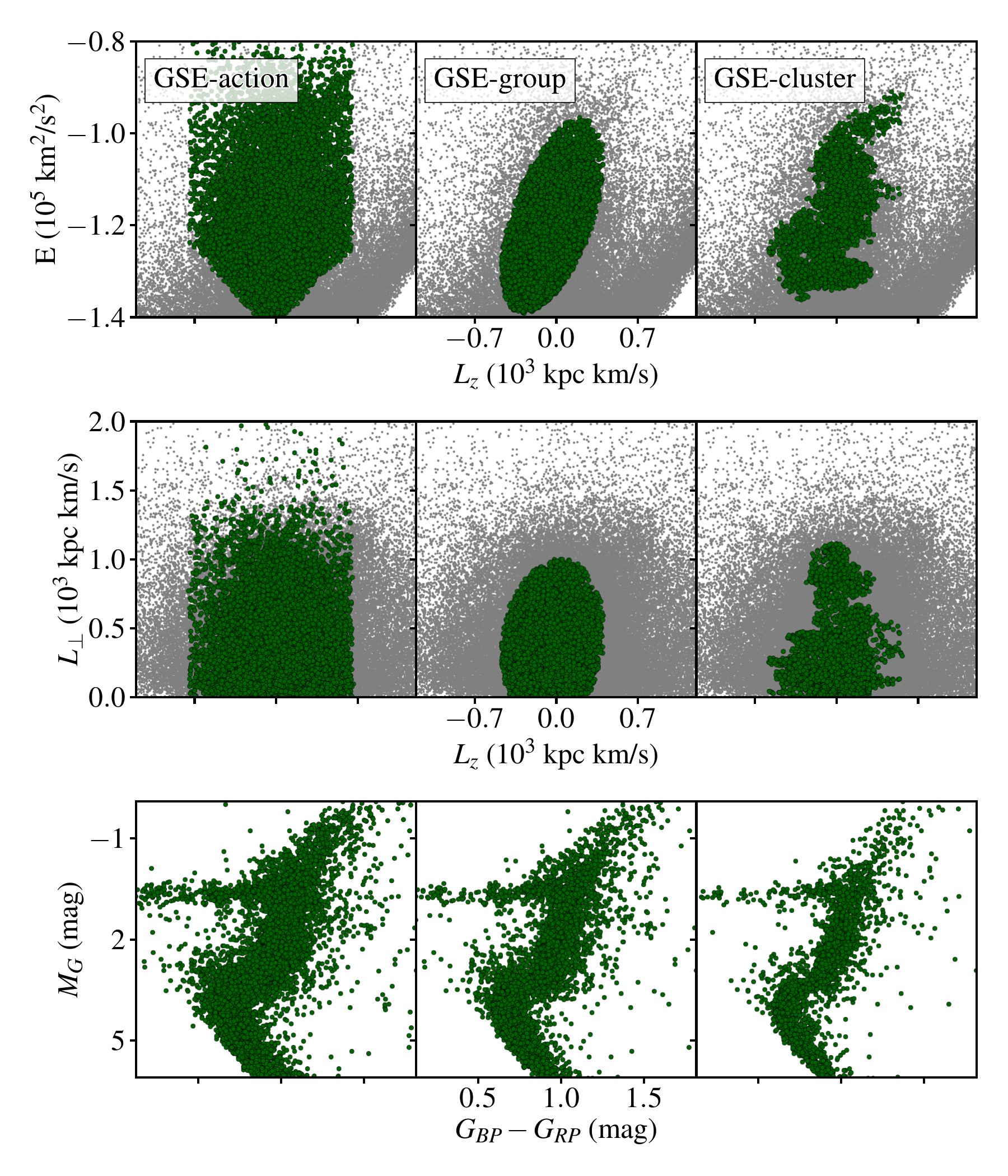}
    \caption{Selection of GSE stellar samples using different criteria. In the top and central panel, the dynamical space (energy versus angular momentum) is represented, showing the whole halo stars sample in grey, and the GSE samples in green. All dynamical coordinates in this plot are calculated using D23 prescriptions for the MW potential and the position and motion of the Sun, to allow for direct comparison. In the bottom panel, the corresponding CMDs are shown. From left to right in decreasing order of number of stars.}
    \label{fig:level_comparison}
\end{figure}

We also follow H24's methodology to define yet another GSE sample, this time
based on action-angle coordinates. For the sake of simplicity, we will reuse 
D23's halo selection as parent sample (D23 sample without quality cuts). Since the action-angle criteria should already
select halo-like stars, the usage of this parent sample is focused solely on making the dataset manageable and consistent with previous samples. After this, and to allow for a direct comparison with H24, we adopt their MW parameters to compute
the action-angle coordinates, namely,
$R_\odot=8.275$ kpc, $z_\odot=20.8$ pc and $(U_\odot, V_\odot, W_\odot)= (8.4, 251, 8.4)$ km/s (see H24 for more information). The potential used is the Gala package~\citep{Price-Whelan2017} \texttt{MilkyWay2022}. The action-angle coordinates $(J_R, J_\phi=L_z, J_z)$ are computed with the AGAMA code~\citep{Vasiliev2019} using the ``Staeckel Fudge'' approximation~\citep{Binney2012}.

The sample is defined by imposing $|L_z|<700$ kpc$\cdot$km/s and a cut in $\sqrt{J_R}$.
Instead of H24's cut, $\sqrt{J_R} > 30$ (kpc$\cdot$km/s)$^{1/2}$ we adopt~\citet{Feuillet2020} selection, $\sqrt{J_R} \in (30, 55) $ (kpc$\cdot$km/s)$^{1/2}$, to avoid extreme radial orbits that do not appear in H24's sample given its limited size (3 out of the 120 stars in H24's GSE sample). This upper limit affects only a small proportion of stars, reducing the sample from 13218 to 12918 stars. Note that GSE-action only includes 75 of H24's stars as a consequence of, mainly, their use of LAMOST radial velocities instead of \textit{Gaia} (\textit{Gaia} DR3 radial velocities are available only for 84 of the stars)\footnote{Even if we partially bridge the
difference between the parent samples, the resulting 
action-angle values for the shared stars slightly differ because of the different distances and radial velocities used.}. 

As a direct reflection of the less restricted dynamical selection, 
GSE-action is the most populated sample 
of the three and contains 90\% and 93\% percent of GSE-group and GSE-cluster. 
We expect the added stars to be a mixture of GSE and other halo structures, making it possibly the most ``contaminated'' sample. The top panel of Fig.~\ref{fig:level_comparison} shows a graphical representation of the samples in integrals of
motion space. Overall, by defining these three samples, we expect to
reduce the effect of the GSE selection criteria in our conclusions, 
as there will be varying degrees of sample ``purity'' (how many 
of the stars belong to GSE) and ``completeness'' (how many of all GSE stars are in the sample).

The following step is to obtain the magnitudes and colours of the stars in the sample, necessary for our CMD fitting. We calculate the absolute magnitude, $M_G$, and colour, $G_{BP}-G_{RP}$, with the zero-point-corrected parallax, and with the reddening and extinction corrections given by the maps from~\citet[][\tt l22]{Lallement2022} and~\citet{Fitzpatrick2019} recipe respectively.
The corresponding CMDs of the three samples can be found in the bottom panels of Fig.~\ref{fig:level_comparison}.
As a consistency check, in Appendix~\ref{app:bayestar} we compare our results following the exact same methodology but with~\citet[][\tt bayestar]{Green2019} reddening maps\footnote{We use the \texttt{bayestar} python implementation~\citep{Green2018}.}. The main
text discussion will focus on the \texttt{l22} case since it allows us
to define more populated samples than \texttt{bayestar}, the latter only covering about three quarters of the sky. The number of stars,
as we will see (Appendix~\ref{sec:AppendixMocks}), is one of the biggest limiting factors when dealing with
our samples in the context of CMD fitting, particularly in this range of age and metallicity.

\begin{figure*}[htpb]
    \centering
    \includegraphics[width=\textwidth]{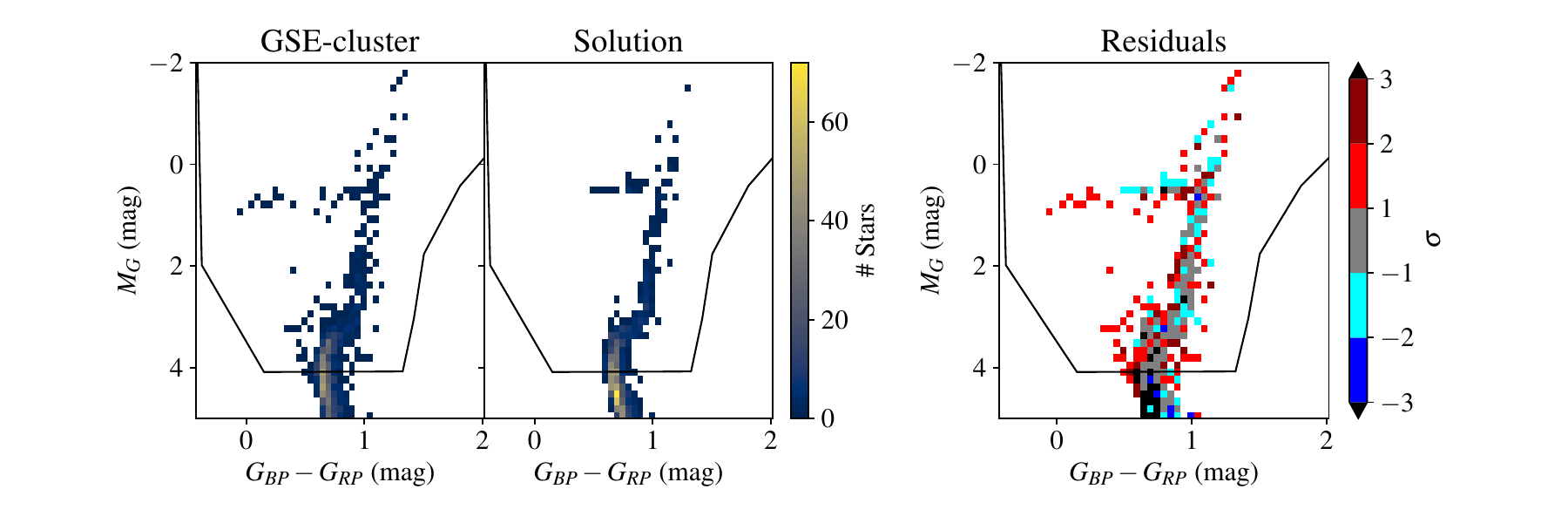}
    \caption{CMD of GSE-cluster sample and corresponding fit and
    residuals. Left panels: quality-cut GSE-cluster sample and corresponding solution CMD of our best-fit deSFH. Only the stars
    inside the region defined by the black solid line are 
    considered for the fit. Rightmost panel: Poissonian
    residuals as $\sigma=(N_\text{observed} - N_\text{solution})/\sqrt{(N_\text{observed}+ N_\text{solution})/2}$.}
    \label{fig:residual}
\end{figure*}

We define now the criteria and quality cuts that will be imposed for the SFH calculation as described in Sec.~\ref{subsec:Method_SFH} and applied
to all samples: 
(i) distance cut, $d<1.2$ kpc\footnote{This distance cut is introduced to ensure high completeness for stars with 6D information at the magnitude of the oldest Main Sequence Turnoff (oMSTO).};
(ii) parallax (with applied zero point correction, \citealt{Lindegren2021}) over error (poe), larger than 5, i.e,
\begin{equation}
    \frac{\varpi}{\sqrt{\sigma_\text{parallax}^2 + \sigma_\text{sys}^2}} \geq 5,
\end{equation}
with $\sigma_\text{parallax}$ being \texttt{parallax\_error} and $\sigma_\text{sys}=0.015$ the uncertainty for the zero-point correction as given by~\cite{Lindegren2021}; (iii) radial velocity error cut, $\sigma(V_\text{los}) < 20$ km/s, (iv) extinction cut, $A_G<0.5$, and (v) cut in excess colour for quality purposes, \begin{equation*} 0.001+0.039\times\text{\texttt{bp\_rp}} < \log_{10}(\text{\texttt{phot\_bp\_rp\_excess\_factor}}),
\end{equation*}
\text{and}
\begin{equation*}
\log_{10}\left(\text{\texttt{phot\_bp\_rp\_excess\_factor}}\right) < 0.12 + 0.039\times\text{\texttt{bp\_rp}}. 
\end{equation*}

Contrary to the definition in D23, we will avoid a \texttt{ruwe} cut, since it systematically removes unresolved binary stars, which are included in the pool of synthetic stars used to fit the CMD.

\subsection{Star formation history calculation}\label{subsec:Method_SFH}
This work uses the CMDft.\textit{Gaia} set of routines which has been described in detail in  ~\citet{Gallart2024}. We will use the terminology introduced in that article. We have used ChronoSynth to compute a synthetic mother CMD that contains 120 million stars with $M_G \le 5$ born with a constant probability for all ages and metallicities (Z) in the range 0.02--13.5 Gyr and 0.0001--0.039, respectively. It has been computed using the updated BaSTI-IAC solar-scaled stellar evolution models~\citep{Hidalgo2018} with a Reimers mass-loss parameter $\eta=0.3$, a fraction of unresolved binary stars $\beta=0.3$, and a minimum mass ratio for binaries $q_{min}=0.1$. The solar-scaled models are used in order
to avoid any prior assumptions about the average $\alpha$-enhancement, $[\alpha$/Fe]. In fact, alpha-enhanced models have been proven to be 
equivalent in optical bands to solar-scaled models under the same global metallicity [M/H]~\citep{Salaris1993,Cassisi2004}. Therefore, when 
mentioning the metallicity, we always refer to the global metallicity, [M/H].

The observed sample without quality
cuts is used as input to the DisPar-\textit{Gaia} algorithm (see Fernandez-Alvar et al., subm.) in order to simulate uncertainties and completeness to this synthetic CMD (dispersed mother CMD). The radial velocity completeness function is implemented as a mask in colour-magnitude-spatial position space that estimates the fraction of stars with radial velocity across the CMD as the ratio between the 6D subsample of {\it Gaia} DR3 and all \textit{Gaia} observed stars (see Fernandez-Alvar et al., subm. for more information).

%

Finally, the dynamically evolved SFH (deSFH\footnote{Similar to \citet{Gallart2024}, we define the deSFH as the amount of mass transformed into stars, as a function of time and metallicity, necessary to account for the population of stars present in a given CMD, in this case the CMD of stars associated to GSE, and present in the studied volume.}) calculation is performed using the DirSFH routine, which finds the best linear combination of simple stellar populations (SSPs) extracted from the dispersed mother CMD that results in the best fit to the observed CMD. We have applied the small offsets of -0.035 mag and 0.04 mag in ($BP-RP$) color and $M_{G}$ magnitude, respectively, between the \textit{Gaia} photometric system and the BaSTI-IAC theoretical framework in the same photometric bands,\footnote{As discussed in \citet{Gallart2024}, uncertainties in the transformation of the stellar evolution predictions from the H-R diagram to the observational plane, as well as residual uncertainties in the photometric calibration, may lead to slight overall systematic shifts between the observed and the synthetic CMDs. In order to derive the size and direction of these possible offsets, an specific routine within DirSFH computes several SFHs with the mother CMD shifted in colour-magnitude space by intervals of hundredths of magnitude. Then, the residuals of these fits are analysed, and an appropriate weighted average of the colour and magnitude shifts leading to the smallest residuals is adopted as the best shift for the final SFH calculation. In \citet{Gallart2024} an excellent \textit{Gaia} stellar sample (that of the stars within 100 pc of the Sun, which is affected by very small photometric errors and negligible reddening, i.e. the Gaia Catalogue of Nearby Stars \citealt{gcns2021}) has been used to estimate a ``generic'' shift between the BaSTI models and the {\it Gaia} observations which, if the distance scale is kept consistent, should be valid for similarly selected \textit{Gaia} samples.} and used the S age bins, and 0.1 dex metallicity bins \citep[see][for details]{Gallart2024}. The fit is limited to a region in the CMD roughly given by $M_G<4.1$; the number of stars inside said region amounts to $N_\text{fit}= 2368, 1339$ and 616 for the GSE-action, GSE-group, and GSE-cluster samples, respectively. The observed CMD, resulting solution CMD, region for the fit, and residuals for the GSE-cluster sample are presented in Fig.~\ref{fig:residual}. 

The distribution of stars in the solution CMD is mostly within $\pm 2 \sigma$ of that in the observed CMD. Note that the solution's horizontal branch (HB) is slightly  brighter than the observed one and less extended to the blue. As discussed in~\citet{Dodd2024} (where the same phenomenon 
is observed when applying CMD fitting to a 5D-defined GSE sample), HB positions depend on a number of 
parameters in addition to age and metallicity (such as He content, see~\citealt{Gratton2010}), while the blue extension depends also on the mass-loss occurred along the RGB phase.  These parameters are fixed in the stellar evolution models used to computed our mother CMD and affect the details of the theoretical HB. However, since the number of stars
in the HB is low, the effect of this discrepancy in our age and metallicity determinations is low. 

Interestingly, there are some bright stars (above the main turn-off) that are not captured by the fit in GSE-cluster (nor GSE-group, and only partially in GSE-action). They resemble a red-coloured main sequence and subgiant-branch in the right-hand panel of Fig.~\ref{fig:residual} (residuals). Some of them may be 
blue stragglers, which are not modelled in our synthetic CMD. All these factors and their low
count makes their fit difficult.

\section{Results and discussion}~\label{sec:results}

We present here the first 6D-based GSE selection age and metallicity characterisation with a CMD fitting methodology. Even if promising, our results are affected by the limitations of the current data, mainly, the low number of stars. 
This occurrence makes challenging the characterisation of the GSE populations made in the present work, with many mock tests done to
correctly interpret the inferred deSFHs. Nonetheless, this paves the path towards what will be possible to accomplish in the upcoming {\it Gaia} data releases.

\begin{figure*}[htpb]
    \centering
    \includegraphics[width=\textwidth]{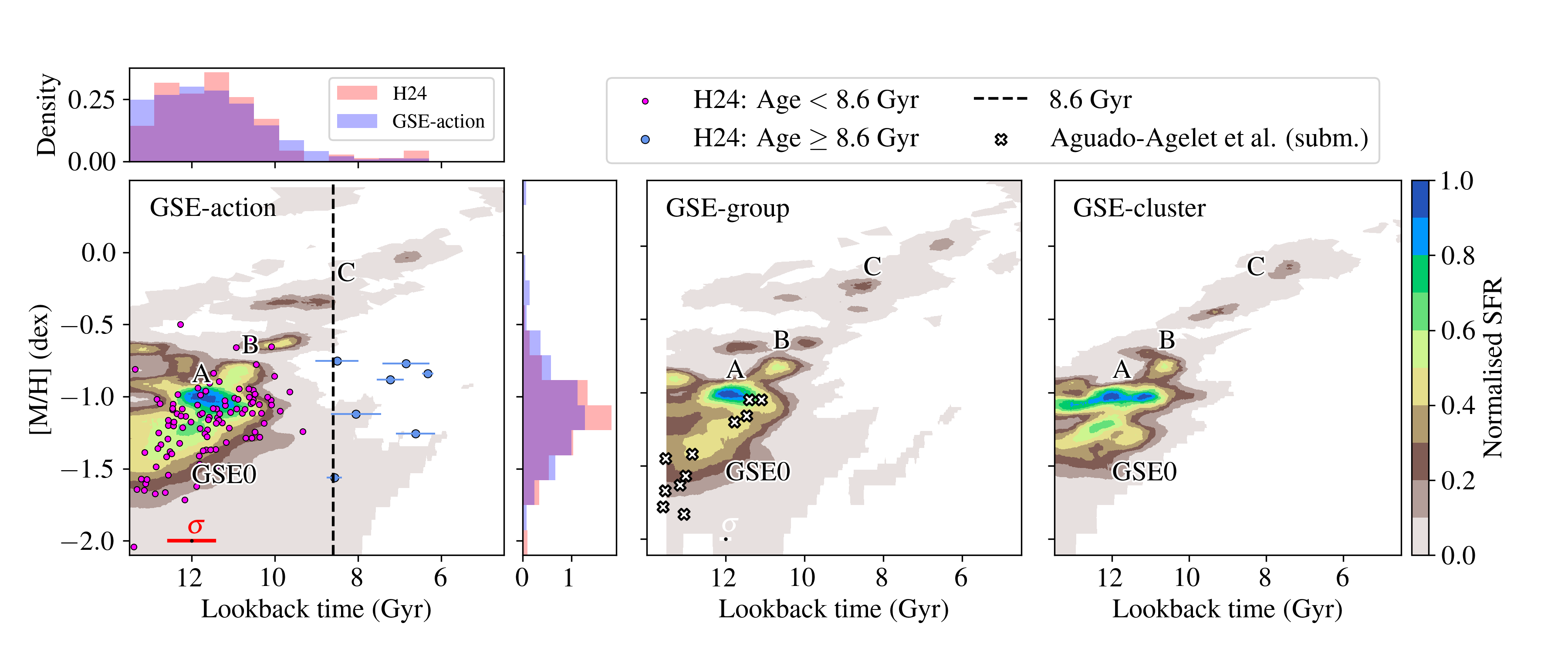}
    \caption{deSFH results for the three samples analysed in this work as filled contour plots of SFR as a function of age and metallicity. In all of them there are four shared features: populations A, B, C, and tail GSE0. In the GSE-action panel, datapoints representing ages and metallicities for individual stars from~\citet{Horta2024} are shown, colour-coded as stars older (magenta) and younger (blue) than 8.6 Gyr; for the younger stars, the age error bars are shown. In the GSE-group panel, ages and metallicities of globular clusters (GC) associated to GSE by Aguado-Agelet et al. (subm.) are represented with white crosses. The graph age range is adjusted in the old limit to accommodate all GCs data (as an age limit of 14 Gyr is adopted for the clusters). In the bottom left corner of the plots with superposed data, data age median uncertainty, $\sigma$, is also shown.}
    \label{fig:GE_three}
\end{figure*}

\begin{figure}
    \centering
    \includegraphics[width=\linewidth]{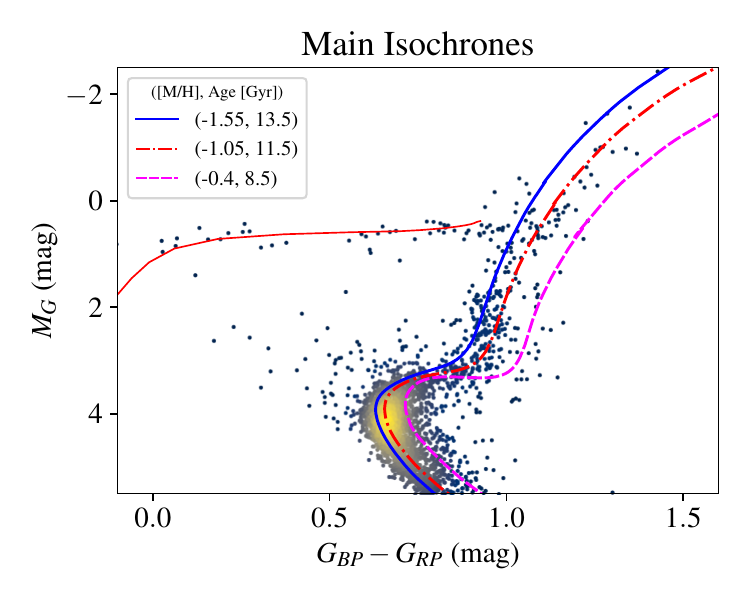}
    \caption{Representative isochrones (up to the tip of the red giant branch) superposed to the observed CMD for the GSE-group (quality-cut) sample. The isochrones have ages and metallicities representative of the oldest population of GSE0, population A and C in the solution, as introduced in Sec.~\ref{sec:results}. Additionally, a zero-age horizontal branch (ZAHB) of metallicity [M/H] = -1.05 with masses in the $M\in [0.55, 0.8]$ $ M_\odot$ range, is presented. We remark that
    the Basti-IAC models are displaced by subtracting to them the 
    offset $(-0.035, 0.04)$ as presented in Sec.~\ref{subsec:Method_SFH}.}
    \label{fig:isochrone}
\end{figure}

In Fig.~\ref{fig:GE_three} we present the normalised dynamically evolved star formation rate (deSFR) as a function of stellar age and metallicity for the three samples of GSE defined in Sec.~\ref{subsec:Method_Data}. We will 
focus our discussion on common features between all three, as it is not 
straightforward to assess
whether the variations come from 
``contamination'' or from methodology-related effects, such as stochastic 
effects introduced by both the selection and the CMD fitting given the small sample size (see also Appendix~\ref{sec:AppendixMocks}).  For a further analysis with \texttt{bayestar} maps and comparison see Appendix~\ref{app:bayestar}. We also note that there is a clear selection bias introduced both by the dynamical and spatial selection (solar neighbourhood). Our GSE selection emphasizes pureness and not completeness~\citep[see also][]{Carrillo2024}, as 6D selections work best when there is a somewhat clear separation between populations. It is likely that GSE is also present in the non-selected lower energy zone~\citep[see][for a low energy age-metallicity distribution]{Dodd2024}.

All panels show three main clumps or populations 
roughly coinciding in age and metallicity and falling in an almost linear age-metallicity relation (AMR). We labelled them from A to C and add to them the feature GSE0, a tail that precedes clump A both in age and metallicity. We will use these four common features, GSE0 and A--C, as reference points to analyse the age-metallicity distribution. The two main features, population A, with (Age [Gyr], [M/H])$\sim$(12$-$11.5, $-1$) and the tail, GSE0, connecting A with the oldest and most metal-poor populations $\sim$(13.5, -1.6) convey the bulk of the star formation activity that took place in the satellite galaxy. As an extension of population A, there are some stars with higher metallicity (B) at 
(10.5, $-0.8$) possibly marking the merger of GSE with the MW, usually inferred to have occurred around 10~Gyr ago~\citep{Gallart2019, Montalban2021NatAs}. Lastly, population C is the youngest and most metallic feature. Even if it comprises few stars: only an average of 4.2\% of stars in the solution CMD are in the associated metallicity range, $-0.5$<[M/H]<$-0.2$ (for \texttt{l22}), we consider it worth discussing since it is present in almost all samples, not only from \texttt{l22} results but also \texttt{bayestar}'s (see Appendix~\ref{app:bayestar}). Feature C could either be driven by the noisy small statistics, indicate a later reignition of the star formation at $\sim$(8.5, $-0.4$), or be a contaminating population, something that will be thoroughly discussed later. In Fig.~\ref{fig:isochrone}, we present 
a CMD with isochrones representative of the mentioned populations.

Note that GSE was originally linked to the blue sequence of the halo CMD reported in~ \citet[][selected kinematically in 5D]{Babusiaux2018}. The first characterisations of its stellar content mainly highlighted its old and metal-poor nature \citep[e.g.][]{Gallart2019}. Recent works, and especially this one, unveil a higher level of complexity and detail when cleaner samples are defined in 6D. 

Given the inherent difficulty due to the small number of stars, we inspect the reliability of the derived populations and of the AMR in Appendix~\ref{sec:AppendixMocks} by means of tests with mock stellar populations\footnote{These tests
are tailored to the~\texttt{bayestar} deSFHs since these 
are more affected by the limitations. However, the conclusions
extracted apply to all samples.}.  From these tests, we can conclude that the general features and AMR are to be trusted. Nonetheless, no distinction can be made between a semi-uniform distribution in the age-metallicity space and a clumpy one when the number of stars is low (especially in GSE-cluster), since both result in the same type of CMD fitting solution. In addition, we also conclude that population A and GSE0 could be systematically  stretched (slightly, $\lesssim$ 10\%) towards older ages and more metal-poor metallicities. This, together with the presence 
of contamination, is probably the reason behind the stretching to older ages present in all plots, 
and situates population A ``actual age'' closer to $\sim 11.5$~Gyr. This possible bias towards older populations was already discussed in~\citet{Gallart2024}.

Excluding population C, the age and metallicity distribution is in line with the values in the current literature, either from stars~\citep[e.g.,][]{Giribaldi2023, Horta2024} or GCs~\citep[e.g.,][]{Massari2019,Forbes2020, Limberg2022,Valenzuela2024}. In Fig.~\ref{fig:GE_three} we directly compare our deSFHs with some state-of-the-art results encompassing both stars and GCs ages and metallicities. In the leftmost panel, stars' ages and total metallicities\footnote{Total metallicities are computed with Eq.1 of~\citet{Gallart2024} using~\citet{Horta2024, Xiang2022} [Fe/H] and [$\alpha$/Fe] values from LAMOST LRS DR7~\citep{Cui2012, Zhao2012}.}  for the sample of GSE stars given by H24~\citep[see][for the age calculation]{Xiang2022} are shown\footnote{The theoretical
models used in~\citet{Xiang2022} are different \citep[Yonsei-Yale,][]{Demarque20024}, and that could have led to different absolute ages, but the relative-age comparison should hold.}. The stars' ages and metallicities are in 
good agreement with our computed deSFH, especially the marginal distributions. The larger differences are our less-peaked metallicity distribution function (MDF) and 
a discrepancy in the youngest stars. Concerning these, on one hand, the not-fitted bright stars, as presented in Sec.~\ref{subsec:Method_SFH} seem partially connected to H24's young population, as $\sim 3$ out of the 7 H24 young stars fall in the 
brighter end above the bulk of the turn-off (as our not-fitted stars). On the other hand, the absence of population C ($\text{[M/H]}=-0.4$) in H24's age-metallicity distribution is plausibly explained by the chemical constraint imposed in H24 ($[\text{Fe/H}]<-0.5$), which limits the pool of stars in the more metallic end. Overall, young populations ($<9$ Gyr) are present in an extensive range of metallicities, even if with a low count.

In the center panel, we show the ages and metallicities for a sample of GCs dynamically associated to GSE~(Aguado-Agelet et al., subm.; see also \citealt{Massari2023})\footnote{The CARMA project relies on the same stellar models library as the Chronogal project (BaSTI-IAC).} estimated from isochrone fitting by the CARMA project. Note that the cluster's AMR tightly follows that of field stars, providing supporting evidence to our results. Additionally, there seems to be two groups of clusters, the first containing very old clusters within a quite broad metallicity range (of nearly 0.5 dex), and a second somewhat younger ($\simeq$11.5 Gyr old) and with a small metallicity range. The position of the older GCs group follows the more metal-poor tail of GSE0, extending further in age and metallicity than our deSFH, while the youngest GCs fall close to population A, some of them being slightly more metal-poor. This distribution of GSE GCs in two groups seems therefore to provide support to the reality of the episodic behaviour of the GSE deSFH, even though some differences are observed:  i) the most metal-poor GCs hint at a high SFR at equal metallicity that is not present in our deSFH and ii) the ages and metallicities of the GCs seem to be slightly displaced to the metal-poor edge of the stellar distribution. Regarding i), note that the aforementioned selection bias may be at play as different GSE populations may be inhomogeneously mixed across the MW. It could be, for example, that the oldest and most metal-poor population is under-represented in our solar neighbourhood sample. The slight displacement can be attributed to the differences between the methodologies and the slightly different distance scales involved in calculating the age and metallicities of GCs and field stars.  In any case, if there is truly a shift, we would expect it to be up to 1 Gyr and $0.2$ dex in age and [M/H], respectively (see Appendix~\ref{app:clusters}). Taking this into account, as a conservative statement, we expect populations A-B to be connected to the youngest star formation epoch (younger GCs) and the oldest GCs 
would fall on the tail GSE0, even if the exact position is not known. 


%
Next, we will discuss plausible hypotheses on merger scenarios compatible with the computed deSFHs and other available observational evidence. If we assume that the oldest stars belonging to clump A are partly the result of age degeneracy, as argued above, we can conclude that the tail, labelled GSE0, could embody the very first star formation of GSE, that is, the primitive GSE, probably before any interaction with the MW. This tail, leading to the youngest population of clump A, is in line with an initial gradual star formation (as opposed to a sudden burst) as postulated in~\cite{Ernandes2024} using [Be/Mg] abundances. The increase in the stellar mass associated with clump A compared to the trailing tail could either reflect the evolution of GSE in isolation or indicate the time for the first gravitational interaction with the MW, $\sim12-11.5$ Gyr ago, which could have triggered enhanced star formation in the satellite galaxy~\citep{Tissera2000, Moreno2019, Cintio2020, orkney2022}.  After that moment, the final significant population (B) is formed, around $\sim 10.5$ Gyr ago (even up to $\sim$ 10 Gyr in GSE-action). This moment in time could coincide with the merging of the GSE progenitor with the MW (estimated to have occurred around $8-10$ Gyr ago based on the youngest ages inferred among GSE debris by e.g. \citealt[][]{Gallart2019, Montalban2021NatAs}; more recently and with precise age estimations: $\sim 9.6\pm0.2$ Gyr,~\citealt[][]{Giribaldi2023}). 

The CARMA GCs ages' bimodality mentioned above (see also middle panel of Fig.~\ref{fig:GE_three}) supports the idea of an episodic star formation, thus reinforcing the idea that there were two main epochs of star formation in GSE: an earlier one, probably in isolation (GSE0), and a younger one (A) connected to the MW pericentric passage and final quenching (after B).
A similar bimodality was also highlighted by~\citet{Valenzuela2024}. In their compilation of GCs ages from several literature sources, these authors found, in some of the analysed samples, a bimodality in the GSE GCs ages, with age separations compatible with the two episodes presented here (12-13 Gyr and 10-11 Gyr). 

%


The first hypothesis regarding the more metal-rich population, with [M/H] $\sim -0.4$ and age $8.5$ Gyr (population C) would be that the progenitor of GSE could have sustained its star formation for 2 Gyr after the usual time inferred for its merger. In this scenario, the youngest and more metal-rich population in C could be the result of a final star formation event in GSE triggered by a further pericentric passage, marking the actual final merger. For this to be the case, the previous pericentric passages must have occurred at a relatively large galactocentric distance to avoid total disruption, and a reservoir of cold gas is needed~\citep{Cintio2020}. This hypothesis was already presented in H24, where they found the 7 mentioned subgiants (shown in blue in the left panel of Fig.~\ref{fig:GE_three}) with ages $\sim 6-9$ Gyr. The possibility for some residual star formation in GSE after the time usually inferred for its merger was also raised in~\citet{Johnson2023}, who applied a one-zone chemical evolution model to interpret the H3 survey data, which finds GSE stars as metal-rich as [Fe/H]=$-0.5$~\citep{Conroy2019, Naidu2021_GES}. Their fit suggests a time-extended star formation (until $\sim 8-9$ Gyr) and reduced SFR after the 10 Gyr mark. This study was already highlighted by~\citet{Ernandes2024}, who linked a [Fe/H]$\sim-0.5$ population with the final quenching of GSE.

The other plausible hypothesis would be that these
stars belong to the in-situ population, as can be seen from the selection in the [Mg/Mn]-[Al/Fe] space in figure~1 of~\citet{Limberg2022}, where the chemical constrain eliminates the more iron-enriched population\footnote{We extract similar conclusions when plotting our [M/H]$>-0.5$ stars in the $[\alpha$/Fe] ([Mg/Fe]) vs [Fe/H] space with LAMOST LRS DR9 (APOGEE DR17,~\citealt{APOGEE}). They fall scattered in the in situ region. }.
Furthermore, dynamical selections are not expected to be completely pure, and this hypothesis is consistent with the expected contamination of the sample, as assessed in~\citet{Carrillo2024}. Further analysis in the topic of the selection procedure would help clarify this point.
Therefore, only upon analysis of larger stellar samples, both 
from spectroscopic surveys and \textit{Gaia} DR4, it will be possible to disentangle the different hypotheses presented here for population C.

\section{Conclusions}\label{sec:conclusions}

In this work, we present the first Gaia-Sausage-Enceladus deSFH, derived using CMD-fitting of stellar samples drawn dynamically using {\it Gaia} proper motions and {\it Gaia} RVS velocities (6D information). We conclude that GSE started forming stars at a smooth rate around 13-13.5 Gyr ago and kept forming stars until 10 Gyr or even 8.5 Gyr ago. Whether this star formation is continuous or episodic is not entirely clear given the limited number of 6D stars belonging to GSE within the studied volume. Nevertheless, comparison with age and metallicity determinations of clusters kinematically associated with GSE (Aguado-Agelet et al., subm.) favours the episodic evolution. As a consequence, we tentatively conclude that GSE had two distinct star formation epochs: 
an older evolution $>12$ Gyr ago, which probably entails the evolution of GSE in isolation, and a second, possibly enhanced, epoch of star formation at $12-11.5$ Gyr probably associated to a pericentric passage. After the 10 Gyr mark, the star formation seems to be mostly quenched.
An extra 8.5 Gyr population of lower significance is also found and discussed. Larger samples would be needed to fully assess its origin.

Despite the limitations, mainly caused by the small number statistics, we have shown the potential of the CMD fitting methodology to unravel the SFHs of the main stellar halo building blocks. The future {\it Gaia} DR4, expected to provide radial velocity for over 100 million sources~\citep{Katz2023} implies an increase of available data that could yield a more robust fit and shed light on the current discussion, ultimately separating the different scenarios that have been presented. Additionally, it would unlock the possible application to smaller structures such as the Helmi streams, Sequoia, Thamnos, etc, nearing a complete knowledge on the accretion history of our Galaxy.

\begin{acknowledgements}
      YGK's work was supported by the ``Beca de iniciación a la investigación'' UGR-Banco Santander grant 2023-2024. TRL acknowledges support from Juan de la Cierva fellowship (IJC2020-043742-I) and Ram\'on y Cajal fellowship (RYC2023-043063-I, financed by MCIU/AEI/10.13039/501100011033 and by the FSE+). 
      
      CG, EFA, SC, AQ, and TRL acknowledge support from the Agencia Estatal de Investigación del Ministerio de Ciencia e Innovación (AEI-MCINN) under grant ``At
      the forefront of Galactic Archaeology: evolution of the luminous and dark matter
      components of the Milky Way and Local Group dwarf galaxies in the Gaia era''
      with references PID2020-118778GB-I00/10.13039/501100011033 and PID2023-150319NB-C21/10.13039/501100011033.
      
      EFA also acknowledges support from HORIZON TMA MSCA Postdoctoral Fellowships Project TEMPOS, number 101066193, 1005 call HORIZON-MSCA-2021-PF-01, by the European Research Executive Agency.
      
      ED and AH acknowledge financial support from a
      Spinoza grant.
        
      SC and DM acknowledge financial support from PRIN-MIUR-22: CHRONOS:
      adjusting the clock(s) to unveil the CHRONO-chemo-dynamical Structure of
      the Galaxy” (PI: S. Cassisi) finanziato dall’Unione Europea – Next Generation EU, and Theory grant INAF 2023 (PI: S. Cassisi). 

      M.M. acknowledges support from the Agencia Estatal de Investigación del Ministerio de Ciencia e Innovación (MCIN/AEI) under the grant "RR Lyrae stars, a lighthouse to distant galaxies and early galaxy evolution" and the European Regional Development Fun (ERDF) with reference PID2021-127042OB-I00, and from the Spanish Ministry of Science and Innovation (MICINN) through the Spanish State Research Agency, under Severo Ochoa Programe 2020-2023 (CEX2019-000920-S).
      
      The authors wish to acknowledge the contribution of the IAC High-Performance Computing support team and hardware facilities to the results of this research. 
      This work has made use of data from the European Space Agency (ESA) mission Gaia (\url{https://www.cosmos.esa.int/gaia}), processed by the Gaia Data Processing and Analysis Consortium (DPAC, \url{https://www.cosmos.esa.int/web/gaia/dpac/consortium}). Funding for the DPAC has been provided by national institutions, in particular the institutions participating in the Gaia Multilateral Agreement.
      
\end{acknowledgements}

%
%
\bibpunct{(}{)}{;}{a}{}{,} 
\bibliographystyle{aa_url} 
\bibliography{biblio}
\begin{appendix}
\section{Alternative reddening map}\label{app:bayestar}

\begin{figure*}[htpb]
    \centering
    \includegraphics[width=\textwidth]{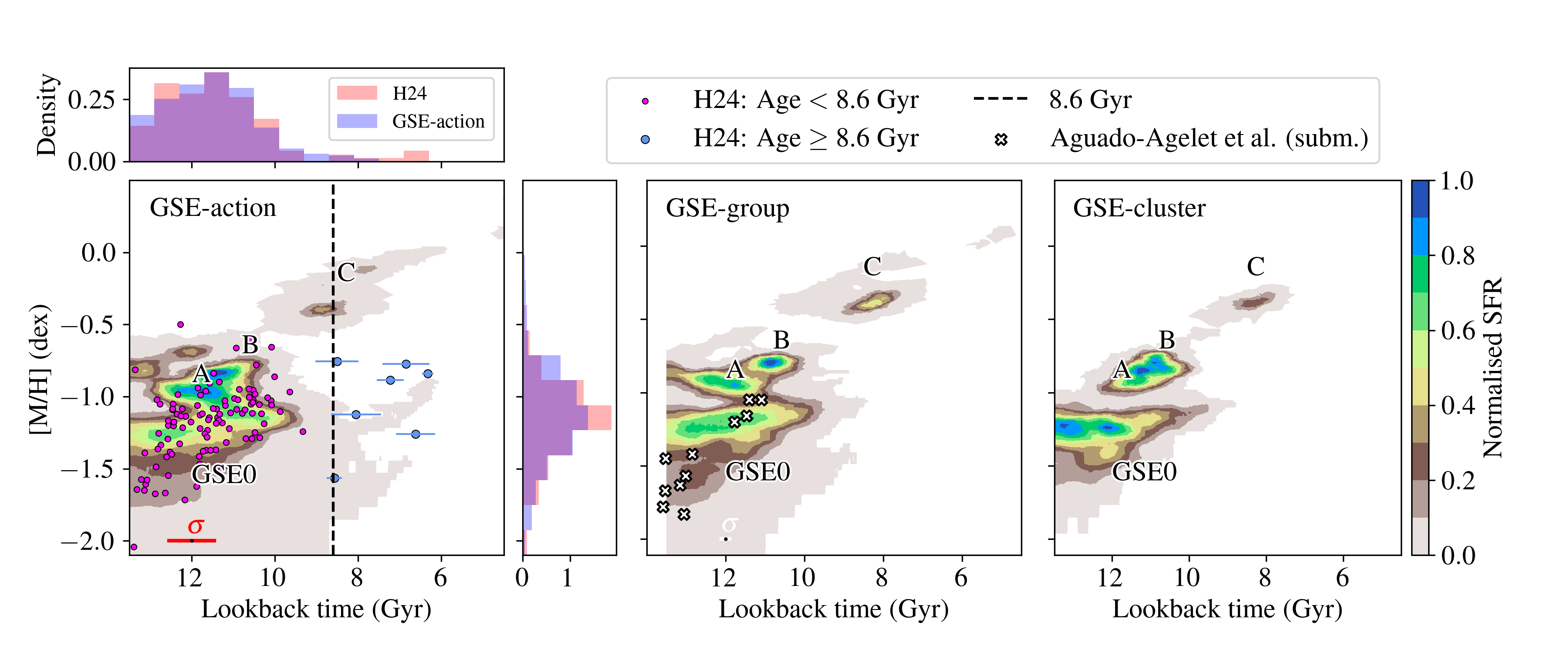}
    \caption{Same as Fig.~\ref{fig:GE_three} but we use \texttt{bayestar} maps for dereddening. Labels' positions and superposed data are the same as in the main text.}
    \label{fig:GE_three_bayes}
\end{figure*}
As introduced in Sec.~\ref{sec:Method}, here we derive and discuss the resulting deSFH when the stars are 
dereddened with the \texttt{bayestar} maps implementation and asses both the similarities and differences with \texttt{l22}. From now on, we will simply refer to them
as \text{l22} and \texttt{bayestar} deSFHs. An exact
copy of Fig.~\ref{fig:GE_three} with the computed \texttt{bayestar} deSFH can be found in Fig.~\ref{fig:GE_three_bayes}. 
We first highlight that the concordance between the main text deSFH and the one presented here increases 
with the number of stars in the sample. In this case, 
the number of stars considered for the fit is  
$N_\text{fit}$= 1748, 973 and 459 for the 
three samples presented in the main text but 
using \texttt{bayestar} to correct for reddening. This is
supporting evidence of how a higher count of stars makes 
the methodology robust. Additionally, many
of the features are shared among all deSFHs, such 
is the case for populations A, B, C, and GSE0, even
if their exact characteristics are not identical.

The dissimilarities between \texttt{l22} and \texttt{bayestar} deSFHs stem  
from the intrinsic differences between the two dust maps 
and are magnified by the CMD fitting algorithm because 
of the likeness of evolutionary tracks with small age and metallicity separation in such old age and metallicity. For example, we checked that the values for the reddening, $E_{BP-RP}$ provided by \texttt{bayestar} are systematically lower than those by \texttt{l22} 
in the approximate range $E_{BP\_RP}\lesssim 0.015$ and tend to be higher than \texttt{l22} for $E_{BP\_RP}\gtrsim 0.015$. The reddening median differences amount to $\sim \pm 0.015$ in both ranges (of the
order of the systematic shift between \textit{Gaia} data and the theoretical models). This is just an example of the complexity of the
comparison between the two and how they can lead to different solution CMDs. 

We believe both results 
derived from \texttt{l22} and \texttt{bayestar} are equally reliable but we give a preference 
to \texttt{l22} solely because of the higher count of stars. With a sufficiently high number of stars, we expect those differences to smooth out, as is the case with GSE-action.

From the features present in \texttt{bayestar} deSFH, the one that stands out the most is the fact that GSE0 appears as a population completely separate from A and B in GSE-cluster. Since this could hint at an episodic star formation similar to CARMA's scenario, we test this hypothesis in Appendix~\ref{sec:AppendixMocks}. We find that, with such 
low number of stars, it is impossible to distinguish 
between two bursts or a semi-uniform star formation. 
To be able to separate between the two scenarios, we need, at least, $N_\text{fit}\gtrsim 1300$ stars.

\section{Mock tests assessing the validity of the results}~\label{sec:AppendixMocks}
Considering that this is the first time that CMDft.\textit{Gaia} has been applied to a population as old and metal-poor as GSE selected in 6D, and the small number of stars in the sample, we have designed some mock populations in order to test the reliability of the results.
In particular, we will focus our attention in the \texttt{bayestar} GSE-cluster sample, as it is the most affected by the low number of stars\footnote{This also means
that for all mocks we will use the \texttt{bayestar} reddening map and samples.}. 
The main idea behind the mock tests is to define
synthetic data with known ages and metallicities 
to assert how they are affected by the
error simulation and DirSFH procedure. By 
studying different mocks we extract information about not only that specific sample, 
but the methodology as a whole.

All of the mock populations will be obtained from another dispersed mother CMD by defining a region in the age-metallicity space from which the synthetic stars are selected. The synthetic population is dispersed
using the (\texttt{bayestar}) GSE-cluster sample. The properties of this second mother CMD are the same as the one used to derive the deSFH of the GSE samples, and described in the main text, except for the metallicity distribution, which in this case is uniform in log$Z$, and for the fact that it contains only 20 million stars. The star count is high enough to guarantee that all mocks can be extracted from a large pool of stars and the uniform age-log$Z$ distribution allows us to build any distribution via appropriate sampling.

\begin{figure}[htpb]
    \centering    \includegraphics[width=0.7\linewidth]{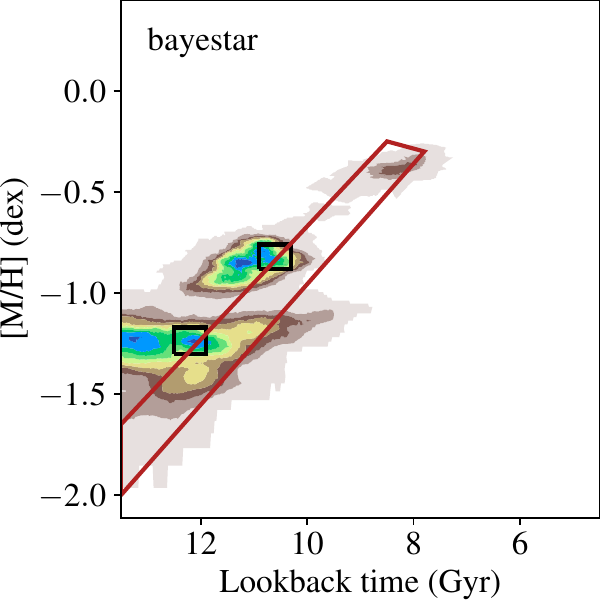}
    \caption{Regions used to
    define the two mocks discussed in the text (``two bursts ''in black and ``uniform'' in red), superimposed on the \texttt{bayestar} GSE-cluster deSFH.}
    \label{fig:GE_mocks}
\end{figure}
As shown in Fig.~\ref{fig:GE_mocks}, two mock populations (``two bursts'' and ``uniform'' star formation) are defined in order to obtain information about how the methodology and characteristics of the data sample, such as a small number of stars, affect the derived GSE deSFH. Both mocks (indicated in the figure with red and black polygons and superimposed on the derived deSFH of the GSE-cluster sample) follow an AMR akin to the one observed in the GSE samples and are tailored to maximise the amount of information that can be extracted from the tests. On one hand, the ``uniform'' mock (red elongated region) is selected from a slightly steeper AMR compared to the one derived for GSE, in order to check whether the AMR and most metal-poor stars ([M/H]$<-1.5$) are correctly retrieved or if there exists any bias towards the observed (flatter) GSE AMR. On the other hand, the ``two burst'' mock focuses on selecting very concentrated populations to check whether the episodes retrieved in the deSFH are more elongated or displaced due to age-metallicity degeneracy. The number of stars drawn is similar to the \texttt{bayestar} GSE-cluster observed sample, $N=1000$\footnote{Down to $M_G=5$, which is the limiting magnitude of the synthetic CMD from which the mock populations are drawn}, resulting in $N_\text{fit}\approx 486$ stars inside the fit region. In the case of the two bursts sample, 600 stars are drawn from the more metal-poor region and 400 from the more metallic one maintaining the proportion between the number of stars in the two separate populations, GSE0 and A-B of the GSE-cluster sample\footnote{Excluding in this case population C altogether.}. Finally, the dispersed mother CMD used for the fit is the one used to derive the GSE-cluster sample deSFH.

\begin{figure}[htpb]
    \centering
    \includegraphics[width=0.5\textwidth]{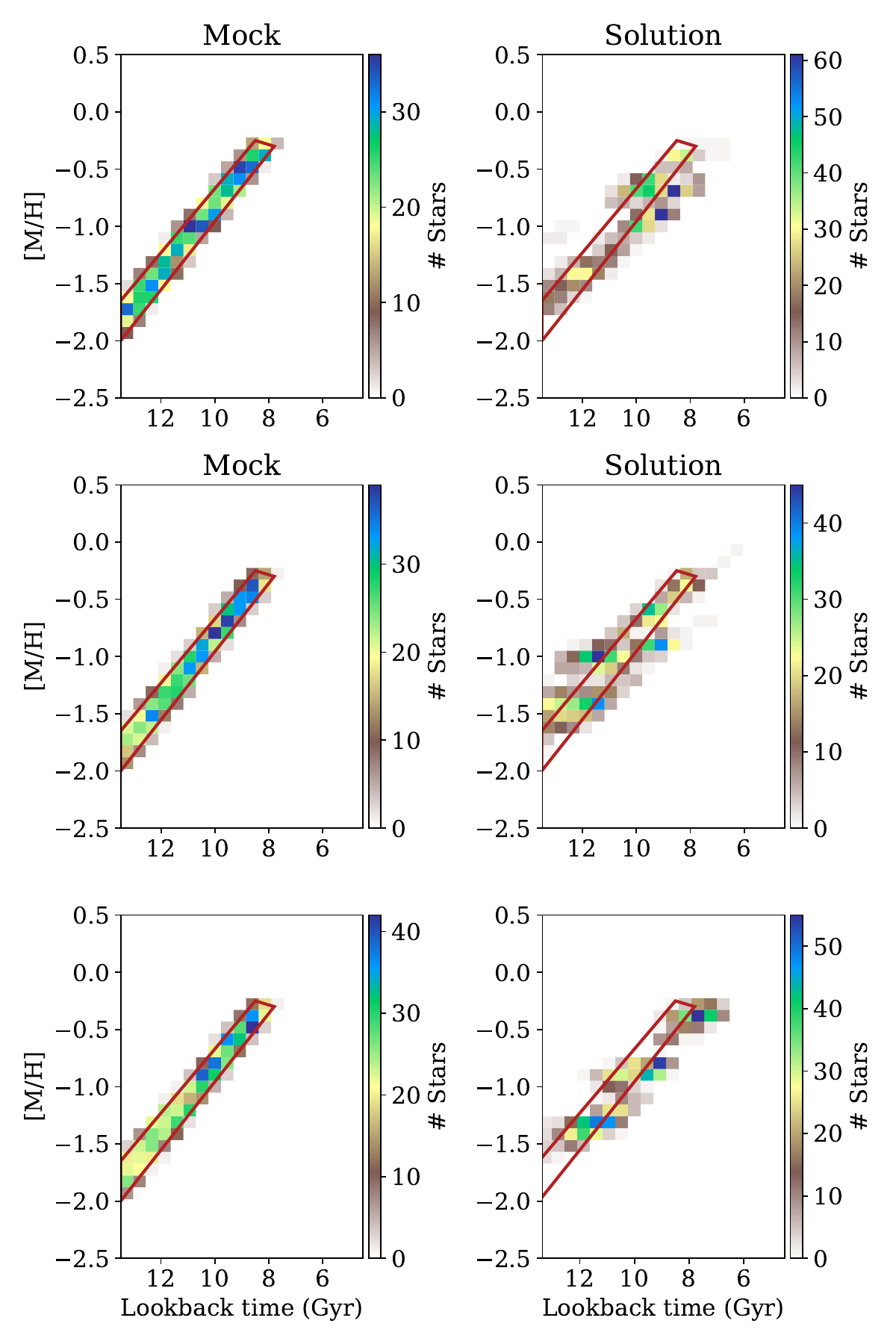}
    \caption{Three different random extractions of $N=1000$, $N_\text{fit}\approx 486$ stars from the ``uniform'' schema. In the left column, the histogram of stars introduced as input. 
    In the right panel, the same plot with the retrieved stars.}  
    \label{fig:GE_mock_seq_n1000}
\end{figure}

    In Fig.~\ref{fig:GE_mock_seq_n1000} we present the histogram of stars in age and metallicity inputted (left) and retrieved in the solution CMD (right) for the ``uniform'' mock. This way, we can compare consistently the inputted distribution and retrieved solution in term of number of stars. We will focus on the conclusions that we would extract based on the plot by visual examination, as in the main text, and make some quantitive statements when necessary.
    The low number of stars in the case of the ``uniform'' mocks results in statistical artefacts that are portrayed as clumps of varying age and metallicity depending on the sample and slight shifts and elongation in the age coordinate ($~\sim 1$ Gyr in the worst cases). The artefacts arise naturally from the statistical variations between random extractions (left column) and are amplified by the methodology. Nonetheless, we can conclude that the AMR (and its slope), and general qualitative features are preserved in the recovered population. Moreover, the oldest and metal-poor stars are (almost)\footnote{The extremely metal-poor population, [M/H]$\sim -1.9$, is never fully retrieved, but it is reasonable given the complexity of the mock. Apart from that, there is no apparent systematic bias in all other ages and metallicities.} as likely to be retrieved as the youngest and metal-rich.

    We draw attention to the fact that the mock random extractions mirror the star-selection procedure followed in our real GSE sample: in our mocks, the underlying distribution
    is known (uniform and following the defined AMR), and the 
    random extraction results in a limited view of the SFH. In a similar manner, GSE-cluster is a subsample of the real GSE population subject to {\it Gaia} incompleteness and our selection criteria (dynamical region, distance, quality cuts, etc). Thus, one could think of the separate bursts in \texttt{bayestar} GSE-cluster, GSE0 and A-B (introduced in
    Sec.~\ref{sec:results} ) as the result of one of such extractions.

    \begin{figure}
        \centering
        \includegraphics[width=\linewidth]{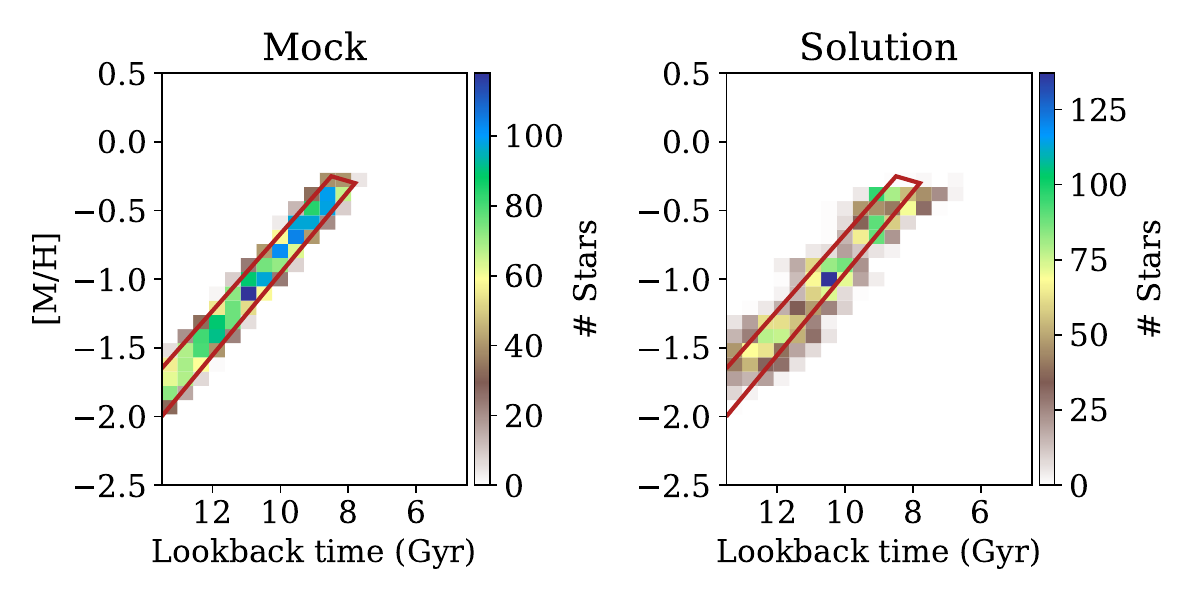}
        \caption{Same as Fig.~\ref{fig:GE_mock_seq_n1000} but 
        with a larger sample, $N=3000$, $N_\text{fit}=1415$.}
        \label{fig:GE_mock_seq_n3000}
    \end{figure}

    We can verify that, in fact, the main factor behind such distortions (especially the elongation and shift in age) is the low number of stars. Fig.~\ref{fig:GE_mock_seq_n3000} shows that the recovered solution for a mock with three times the number of stars is closer to the input mock. With this number of stars, the solution is still clumpier than the input mock, but the qualitative resemblance between the input data and the solution is improved. This test highlights the improvement expected in this kind of analysis with the use of upcoming {\it Gaia} data releases. This also supports the validity of the deSFH in the most populated sample, GSE-action, whether it truly
    contains more ``contamination'' than the other two or not.

\begin{figure}[htpb]
    \centering
    \includegraphics[width=\linewidth]{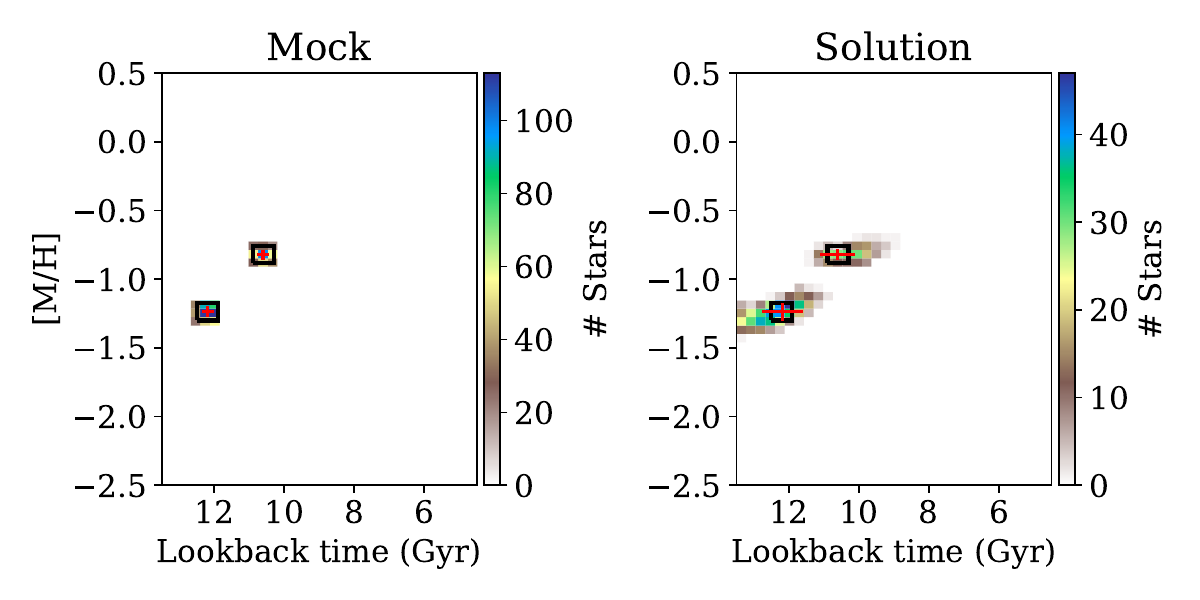}
    \caption{Comparison between ``two burst'' input and the stars retrieved. Same plot as in Fig.~\ref{fig:GE_mock_seq_n1000}. Red crosses indicate the mean and standard deviation for the distributions of each ``burst''.} 
    \label{fig:GE_mock_two_burst}
\end{figure}

    In the case of the ``two bursts'' mock, we can conclude from Fig.~\ref{fig:GE_mock_two_burst} that simple populations 
    such as these are correctly recovered with slight distortions
    in the age and metallicities distributions. Although the number of stars in each burst is only approximately conserved (probably due to some degeneracy in the fit), with 674 and 293 stars in each burst respectively (as opposed to 600 and 400 in the input), if the GSE population is composed of two bursts, we would correctly recover their ages and metallicities with less than 10\% error. For our mock, the age and metallicity means are very closely replicated; the populations (Age [Gyr], [M/H]) inputted as $(12.2, -1.23)$ and $(10.6, -0.81)$ are recovered as $(12.4, -1.24)$ and $(10.3, -0.81)$, but there is a clear increase in dispersion, with the standard deviation in age going from $\sigma_{\text{age}} = 0.18$, $\sigma_{\text{age}}=0.17$ to $\sigma_{\text{age}}=0.57$ and $\sigma_{\text{age}}=0.5$ for the older and younger bursts respectively. The age elongation in the oldest burst can also be used as an argument to support the idea that the actual age of the oldest populations in the GSE deSFHs is in the younger end of their respective estimates.
    
    All in all, although these tests do not cover all possible instances, they advise us on to what extent we can trust all the details found in the DirSFH results and, consequently, on the conclusions we can draw from our analysis.

    
\section{Mock test of an extreme case: fast enrichment}
We present here another mock sample that represents an extreme case of chemical enrichment, in which a chemical build-up is experienced in a very short time window ($\sim 1$ Gyr). This test is meant to further showcase the limitations of the methodology at this point in time for small stellar samples.
\begin{figure}
    \centering
    \includegraphics[width=\linewidth]{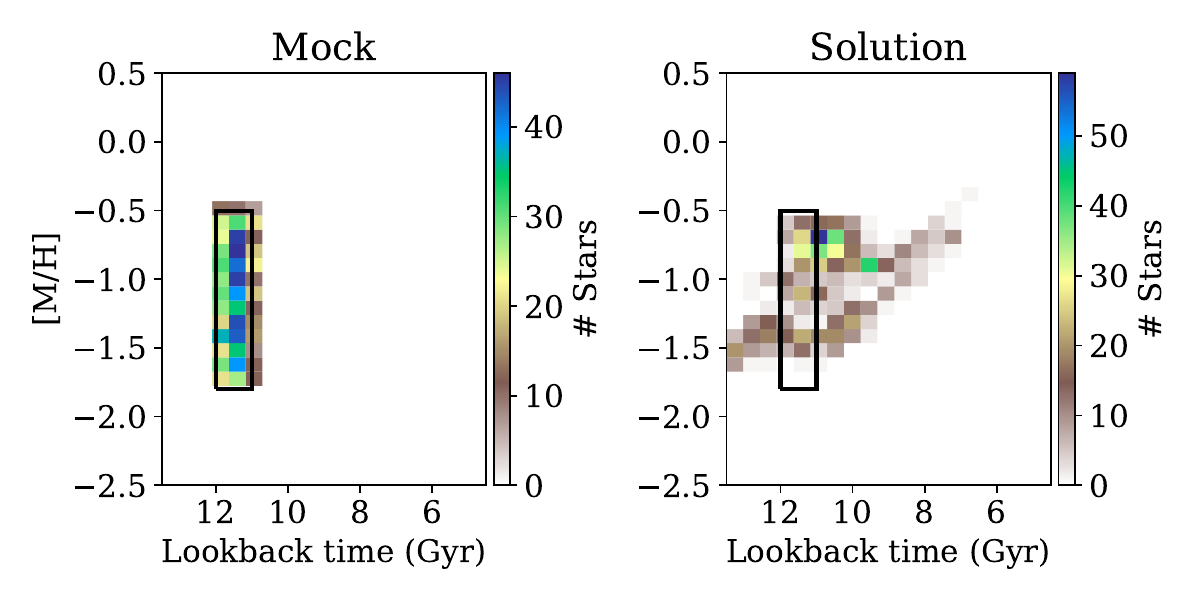}
    \caption{Same as Fig.~\ref{fig:GE_mock_seq_n1000} with the fast-enrichment mock.}
    \label{fig:GE_mock_seq_vert_n1000}
\end{figure}
\begin{figure}
    \centering
    \includegraphics[width=\linewidth]{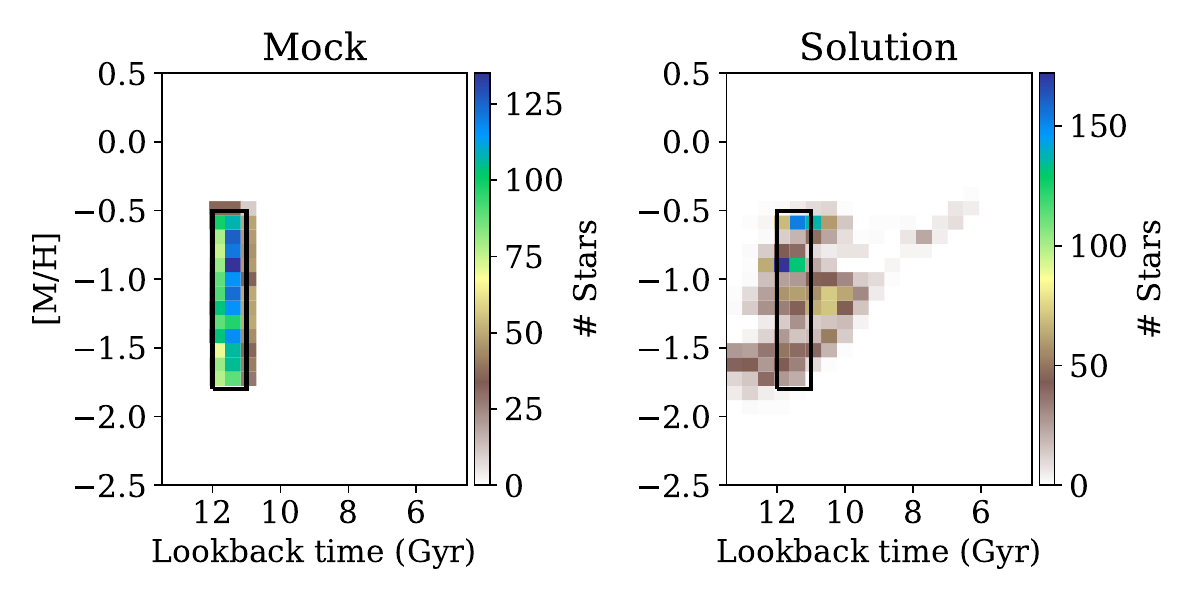}
    \caption{Same as Fig.~\ref{fig:GE_mock_seq_vert_n1000} 
    but with a larger sample of $N=3000$ stars.}
    \label{fig:GE_mock_seq_vert_n3000}
\end{figure}

The methodology followed is the same as in Appendix~\ref{sec:AppendixMocks} with 
the defined region, mock, and solution shown in 
Figs.~\ref{fig:GE_mock_seq_vert_n1000} and~\ref{fig:GE_mock_seq_vert_n3000}, which show samples
drawn with $N=1000$ ($N_\text{fit}=475$) and $N=3000$ ($N_\text{fit}=1399$) respectively. Firstly, when the number of stars is very low (similar to our GSE-cluster observed samples, Fig.~\ref{fig:GE_mock_seq_vert_n1000}) some 
artifacts appear as a direct result of the physical (and uncertainty-induced) age-metallicity degeneracy. As such, 
the retrieved solution is far from the mock input, with a slope 
for the AMR slightly closer to the one derived from the GSE sample. Because of this, strictly speaking we cannot discard the possibility that our observed GSE deSFH is affected by this age-metallicity degeneracy, resulting in a retrieved AMR less
steep than the real one. 

The resemblance between the input mock and the solution 
is much tighter for a larger sample (in this case $N_\text{fit}=1399$) with only a residual population following an AMR similar to the one followed by the derived GSE SFH. Since our GSE-action samples have 
over that number of stars and no features similar to our fast enrichment scenario are present, this possibility seems very unlikely. Other disregarding arguments are the concordance between 
other works in the literature and our solution, and the well-known GSE spread in $[\alpha/$Fe]~\citep[e.g.,][]{Helmi2018} suggesting a spread in age.

\section{Testing CMDft.Gaia on globular clusters}
\label{app:clusters}

As discussed in Sec.~\ref{sec:results}, the AMRs of the GSE dwarf galaxy, derived with either field stars (this work) and GCs (Aguado-Agelet et al., subm.), follow similar trends and show at least two episodes of star formation. Given the drastically different nature of both analysis, understanding how well both results can be compared to each other is of the utmost importance, not only to further assess the reliability of the results, but to properly unveil the past evolution of GSE. A possible small systematic offset in the age-metallicity plane is observed between the two, with GCs appearing near the metal poor edge of the field stars distribution, especially in the case of the l22 solutions discussed in the main text. To address whether this shift is real or arises from systematic differences steaming from the use of different methodologies and photometric systems, we derived the age and metallicity of two GSE GCs using CMDft.Gaia and \textit{Gaia} data. 

To define the sample of GCs member stars we used the catalogues provided by \citet{vasiliev&baumgardt2021}, selecting only stars with a probability membership $> 90 \%$. Moreover, we focused our analysis only on the external regions of the GCs by removing all member stars within $1\sigma$ of their spatial distribution. In this way we exclude the most crowded region, where \textit{Gaia} photometry may suffer from bad measurements, and reduce the impact of the multiple populations phenomenon in GCs~\citep[][]{cassisi&salaris2020}, because second generation stars are commonly more concentrate than first generation ones. Among the list of GSE GCs analysed in Aguado-Agelet et al. (subm.), we selected only the two most populated ones, as they will ensure the most robust results. One is part of the older and more metal-poor group, which could be associated to GSE0 (i.e. NGC6341) and the other belongs to the younger and more metal-rich group, closer to clump A (i.e. NGC362). 

The analysis that, for simplicity we will call "SFH calculation"\footnote{The SFH of a GC cannot be realistically derived with CMDft.\textit{Gaia}, since the range of ages and metallicities of GCs are narrower than those of the SSPs that we use in the analysis.  Thus a perfect fit is basically impossible if we want to keep the conditions identical to the SFH derivation of the field stars. However, this exercise can still be used to estimate the precision of the age and metallicity measurements of CMDft.\textit{Gaia} in this regime of ages and metallicities, as well as possible systematic effects due to the different data and methodology.} has been performed as outlined in Sec.~\ref{subsec:Method_SFH}. However, since GCs are significantly more distant and crowded compared to field stellar populations, the inversion of \textit{Gaia} parallax is not the best proxy to estimate their distance. Therefore, the procedure followed to calculate the absolute magnitudes of member stars has been somewhat different than in the case of the field stars described in Sec~\ref{sec:Method}. Specifically, we assumed that all stars within a given GC are located at the same distance from the Sun and suffer the same extinction. We explored two sets of solutions using different values for distance and reddening. First, we adopted the mean distance estimates by \citet{baumgardt&vasiliev2021} and the reddening provided by \citet{harris2010}. Second, we employed the distance and reddening parameters derived from the analysis in Aguado-Agelet et al. (subm.) within the CARMA project~\citep{Massari2023}. The latter approach ensures that any differences in the clusters' age and metallicity estimates stem solely from variations in methodology and photometric data. We will refer to them, respectively, as the `BV solution' and the `AA solution' from now on. Among the quality cuts described in Sec.~\ref{subsec:Method_Data}, we only applied those in brightness and \texttt{phot\_bp\_rp\_excess\_factor}. The fit is performed in the same region of the CMD defined in Sec. \ref{subsec:Method_SFH}. Thus, the final samples comprise 1400+ and 1200+ member stars inside the bundle for NGC 362 and NGC 6341, respectively. Since we did not use \textit{Gaia} parallax to estimate the distance of GCs, we didn't simulate observational errors derived from the errors of the \textit{Gaia} parallaxes in the synthetic CMD. To account for potential systematic offsets between theoretical models and absolute \textit{Gaia} CMDs, we derived SFHs using the fiducial shift in colour and magnitude of (-0.035, 0.040) derived in \citet{Gallart2024} for field stars using the corresponding DirSFH routine.
%
\begin{table*}
\caption{Main properties of the analysed GCs.}          
\label{tab:GCs}      
\centering          
\begin{tabular}{cccccccc}  
\hline      
GC & Solution & $\mathrm{[M/H]}_{\mathrm{CARMA}}$ & $\mathrm{Age}_{\mathrm{CARMA}}$ & $\Delta$ $\mathrm{[M/H]}_{13.5}$ & $\Delta$ $\mathrm{Age}_{13.5}$ & $\Delta$ $\mathrm{[M/H]}_{14.0}$ & $\Delta$ $\mathrm{Age}_{14.0}$  \\ 
& & (dex) & (Gyr) & (dex) & (Gyr) & (dex) & (Gyr)   \\ 
\hline \\
\vspace{0.2cm}
NGC 362	 & 	 BV	 & 	-1.21 \(^{+0.02}_{-0.02}\)	 & 	11.46 \(^{+0.10}_{-0.09}\)	 & -0.11 & 0.79 & 	0.19	 & 	0.53	 \\
\vspace{0.2cm}
NGC 362	 & 	 AA  &  -1.21 \(^{+0.02}_{-0.02}\)	 & 	11.46 \(^{+0.10}_{-0.09}\)	 & -0.06 & 0.07 & 0.04 & 0.09 \\
\hline \\  
\vspace{0.2cm}
NGC 6341	 & 	 BV	 & 	-1.77 \(^{+0.08}_{-0.09}\)	 & 	13.58 \(^{+0.22}_{-0.21}\)	 & 0.22 & 1.01 & 	0.22	 & 	0.79 \\
 \vspace{0.2cm}
NGC 6341	 & 	 AA  &  -1.77 \(^{+0.08}_{-0.09}\)	 & 	13.58 \(^{+0.22}_{-0.21}\)	 & -0.02 & 0.61 &  0.01 & 0.40 \\
\hline   
\end{tabular}
\tablefoot{We list the values of the age and metallicity derived from the isochrone fitting ($X_{\mathrm{CARMA}}$) in Aguado-Agelet et al. (subm.) together with the differences with median values derived from SFH for both solutions using the mother CMDs with SSPs as old as 13.5 and 14.0 Gyr (e.g., $\Delta X_{13.5} = X_{\mathrm{CARMA}} - X_{13.5}$).}
\end{table*}

\begin{figure*}[htpb]
   \centering
        \includegraphics[width=\textwidth]{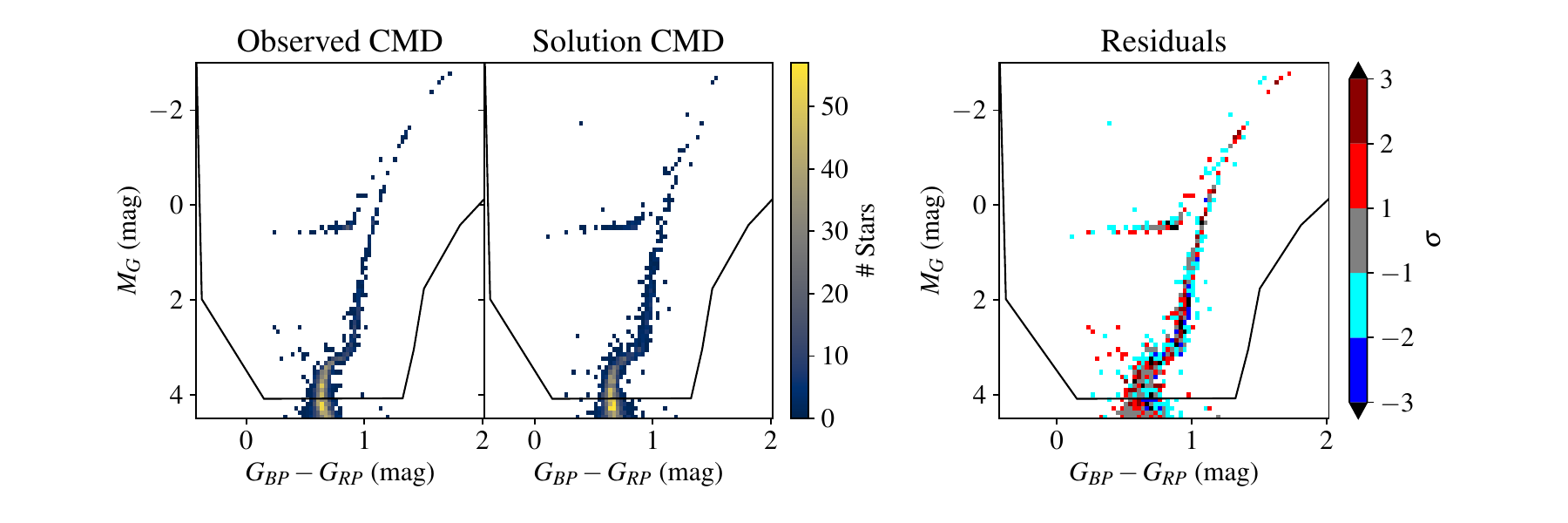}     
   \caption{Same as Fig.~\ref{fig:residual} but for the GC NGC 362 (`AA' solution).}
              \label{fig:CMD_NGC362_CARMA}%
\end{figure*}

In Fig.~\ref{fig:CMD_NGC362_CARMA} we show the \textit{Gaia} CMD of NGC 362 transformed to the absolute plane using the distance and reddening from Aguado-Agelet et al. (subm.), the best `AA solution' CMD found by DirSFH and the residuals between the two. The characteristics of the CMD sequences are well recovered, including the extent of the horizontal branch, which is quite challenging given the uncertainties in the modelling of this advanced stellar evolution phase and the uncertain parametrisation of the mass-loss along the red giant branch phase. Most residuals are within $\pm 3 \sigma$ and they are mainly associated to the fact that the sequences in the solution CMDs are somewhat thicker than in the observed CMD. This is expected, since a GC has a very small (if any) range of age and metallicity, while the SSPs that are used in the fit are around 500 Myr (at the old age of the cluster) and 0.1 dex wide in age and metallicity, respectively.  

\begin{figure}[htpb]
   \centering
   \begin{minipage}{0.24\textwidth}
        \centering
        \includegraphics[width=1.0\textwidth]{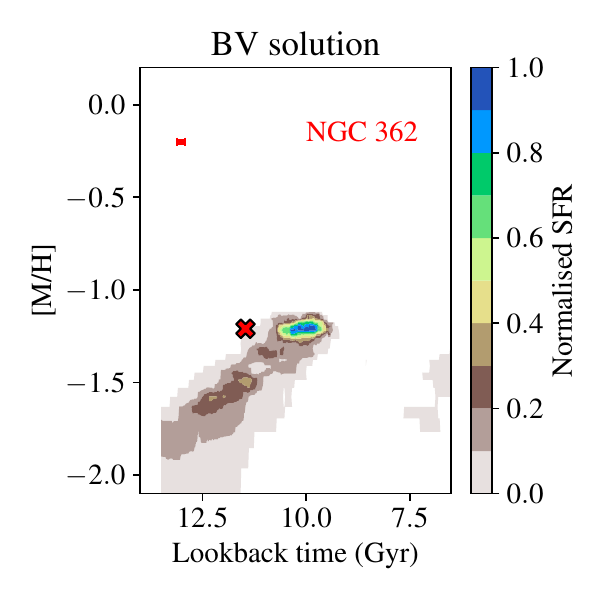} 
        \includegraphics[width=1.0\textwidth]{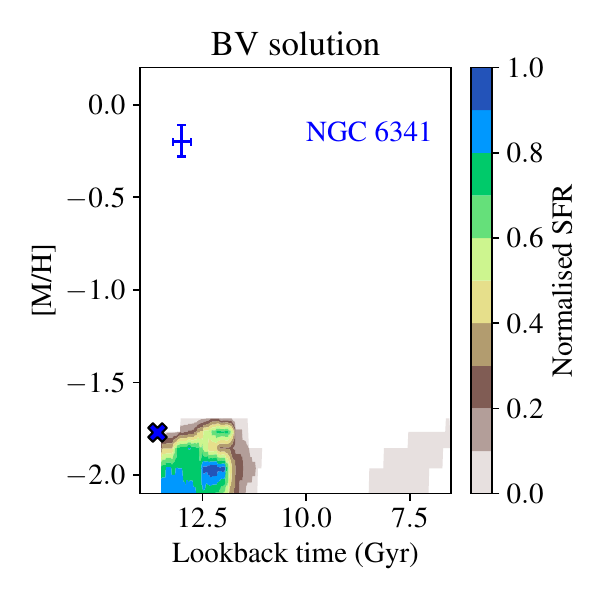}         
   \end{minipage}
   \begin{minipage}{0.24\textwidth}
        \centering
        \includegraphics[width=1.0\textwidth]{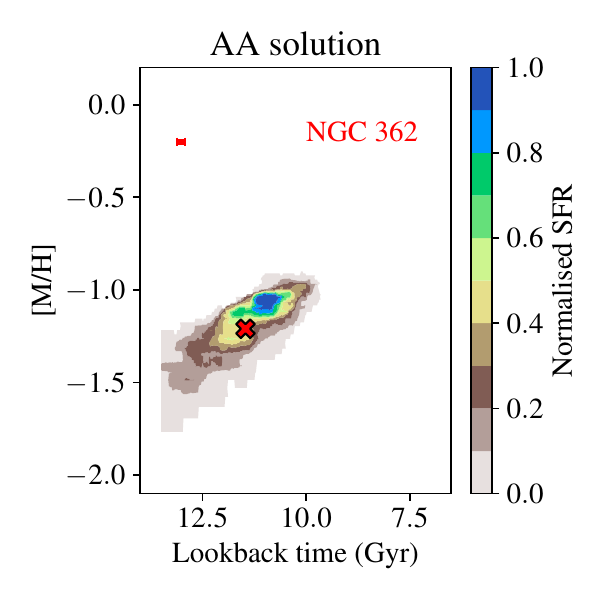}
        \includegraphics[width=1.0\textwidth]{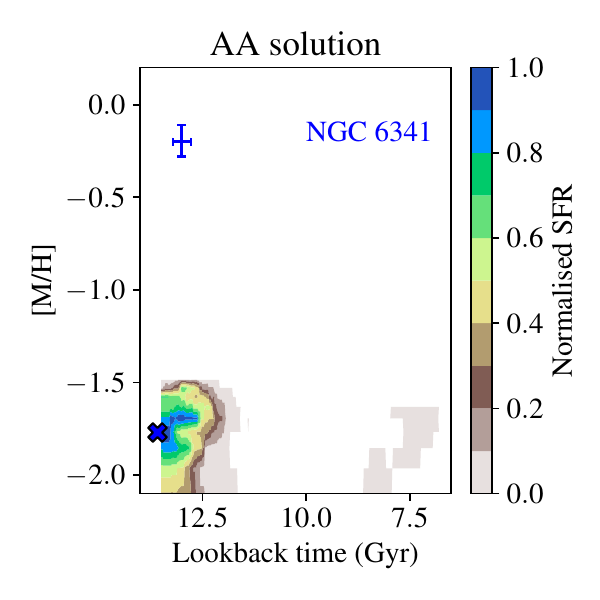}        
   \end{minipage}   
   \caption{SFH obtained for the `BV solution' (left column) and for the `AA solution' (right column) for the GCs NGC 362 (top row) and NGC 6341 (bottom row). The result obtained by Aguado-Agelet et al. (subm.) is reported with a filled cross with the associated uncertainties shown in the top left corner of each panel.}
              \label{fig:GCs_fid}%
\end{figure}

In Fig.~\ref{fig:GCs_fid} we present the results of the SFH calculation for the two GSE GCs for the `BV' and `AA' solutions (left and right columns, respectively). As reference values, we also include the age and metallicity values obtained by Aguado-Agelet et. al (subm.) depicted as red and blue filled cross for NGC 362 and NGC 6341, respectively. For both clusters and across the two solutions, we identify a single prominent burst of star formation, as expected for GCs. However, we also observe an extended metal-poor tail, which is likely attributable to the width in age and metallicity of the SSPs used in the analysis (see discussion above). Also, we note that the faint signal observed in some solutions at ages $<8.5$ Gyr are likely artifacts of the DirSFH fitting process attempting to account for the presence of blue straggler stars in the observed CMD. These stars, which are products of complex stellar evolution, are not explicitly modelled in the code and thus introduce this spurious effect in the solution CMD. 

\begin{figure}
   \centering
   \includegraphics[width=.45\textwidth]{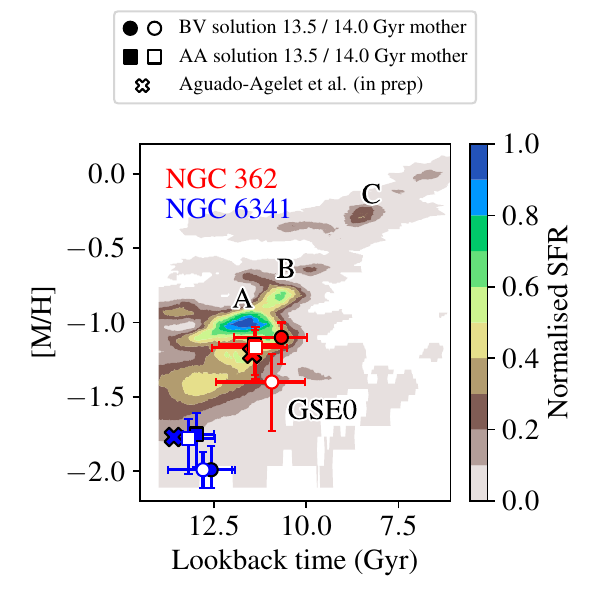}
   \caption{Median age and metallicity in the solution CMD of the analysed GCs obtained with DirSFH. Filled and open symbols represent values obtained with a mother CMD with an upper limit in age of 13.5 and 14.0 Gyr, respectively. Uncertainties are computed at the 16th and 84th percentile of their respective distributions.}
              \label{fig:AMR_SFH_GCs}%
    \end{figure}

The age and metallicity of the two GCs for each of the CMD fits are determined as the median values of the respective distributions of the stars of the solution CMDs, while the associated uncertainties are quantified at the 16th and 84th percentile. The final results for NGC 362 (red filled symbols) and NGC 6341 (blue filled symbols) are plotted in Fig.~\ref{fig:AMR_SFH_GCs}, together with the ages from Aguado-Agelet et al. (subm.) and listed in Table~\ref{tab:GCs}. Open symbols refer to age and metallicity derivations using a mother CMD with maximum age of 14.0 Gyr. In background we plot the solution from DirSFH for the GSE-group sample and the \texttt{l22} reddening map using a mother CMD with maximum age of 14.0 Gyr.\footnote{We take this chance to show that the upper age limit of the synthetic stars in the mother CMD affects very little the features of the solution, as can be seen by comparing this plot with the central panel of Fig.~\ref{fig:GE_three}.} In the `BV' solution, the median metallicity recovered from the solution CMD is either overestimated (NGC 362) or underestimated (NGC 6341) compared to Aguado-Agelet et al. (subm.). Interestingly, the solution for both GCs is somewhat younger by $\sim$ 1 Gyr.  In the `AA' solution, for which we adopt the distance and reddening derived by Aguado Agelet et al. (subm.), we find no shift in metallicity between methodologies. Also, the age of NGC 362 is well-recovered, while in the case of NGC 6341 we find a solution that is 0.60 Gyr younger. These differences in the DirSFH values are mainly due to the differences in the distance and E(B-V) adopted for the two solutions. We note that the difference in age between the ages of Aguado-Agelet et al. (subm.) and our ages may be amplified by our use of a mother CMD with ages in the range of 0.02 - 13.5 Gyr (as described in Sec.~\ref{subsec:Method_SFH}), whereas in Aguado-Agelet et al. (subm.) ages up to 14.0 Gyr are allowed in the isochrone library, and in fact, the age estimate for NGC 6341 exceeds our 13.5 Gyr limit. To evaluate the robustness of our results, we recalculated the SFH for NGC 362 and NGC 6341 using an alternative mother CMD that extends the age range up to 14.0 Gyr. We find that this adjustment has minimal impact on our results, with a mild improvement in the agreement with values from Aguado-Agelet et al. (subm.): the median age and metallicity of the target GCs are, on average, 0.3 Gyr older and 0.1 dex more metal-poor (see open symbols in Fig.~\ref{fig:AMR_SFH_GCs}). 

In Table~\ref{tab:GCs} we summarize the comparison between ages and metallicities derived with the two methods and the different assumptions. We can conclude that, while the differences in the age and metallicity determination with the two different methodologies and datasets are consistent within the errors if the same distance and reddening are adopted, systematic differences of the order of up to 1 Gyr in age and 0.2 dex in metallicity could arise from systematic differences in the distance and reddening scales. Therefore, we cannot exclude a possible relative shift in the ages and metallicities of GCs and field stars that could account for the different position of clumps A and B in the field star SFH and that of the two groups of GCs. Given the inherent complexity of the comparison, 
it is impossible to give a unique absolute shift between both data (CARMA and this work), but the
comparison is possible having this in mind.

\end{appendix}

\end{document}